%% file: main.tex
\documentclass{aa}  

\usepackage{graphicx}
\usepackage{txfonts}
\usepackage[colorlinks=true,linkcolor=blue,citecolor=blue]{hyperref}%
\usepackage{xcolor}
\usepackage{braket}
\usepackage{subcaption}
\usepackage{ctable}

\usepackage{graphicx}	
\usepackage{amsmath}	
\usepackage{comment}

\newcommand{\OIIIa}{[OIII]$\lambda4363\,$}	
\newcommand{\OIII}{[OIII]$\lambda5007\,$}
\newcommand{\OII}{[OII]$\lambda3727\,$}
\newcommand{\OIId}{[OII]$\lambda\lambda3726,3728$}

\newcommand{\NII}{{[N\,{II}]\ }}
\newcommand{\NeIII}{[Ne\,{III}]$\lambda3869\ $}
\newcommand{\SII}{{[S\,{II}]}}

\newcommand{\NV}{{N\,{V}\,}}
\newcommand{\NeIV}{{[Ne\,{IV}]}}

\newcommand{\NeV}{{[Ne\,{V}]}}

\newcommand{\OIIId}{{[O\,{III}]\,$\lambda\lambda4959,5007\,$}}
\newcommand{\HeII}{{He\,{II}\,$\lambda4686\ $}}

\newcommand{\Ha}{H$\alpha\ $}
\newcommand{\Hb}{H$\beta\ $}
\newcommand{\Hg}{H$\gamma\ $}

\begin{document}

   \title{New AGN diagnostic diagrams based on the [OIII]$\lambda 4363$ auroral line}
    \titlerunning{AGN selection based on \OIIIa}
   \subtitle{}

   \author{Giovanni Mazzolari \thanks{\email{giovanni.mazzolari@inaf.it}}
          \inst{1,2,3},
        Hannah Übler,\inst{3,4}
        Roberto Maiolino, \inst{3,4,5}
        Xihan Ji,\inst{3,4}
        Kimihiko Nakajima,\inst{6}
        Anna Feltre,\inst{7}
        Jan Scholtz,\inst{3,4}
        Francesco D'Eugenio,\inst{3,4}
        Mirko Curti,\inst{8}
        Marco Mignoli,\inst{2}
        Alessandro Marconi\inst{7,9}
          }
\authorrunning{G. Mazzolari et al.}
   \institute{  Dipartimento di Fisica e Astronomia, Università di Bologna, Via Gobetti 93/2, I-40129 Bologna, Italy
\and INAF – Osservatorio di Astrofisica e Scienza dello Spazio di Bologna, Via Gobetti 93/3, I-40129 Bologna, Italy
\and Kavli Institute for Cosmology, University of Cambridge, Madingley Road, Cambridge, CB3 0HA, UK
\and Cavendish Laboratory, University of Cambridge, 19 JJ Thomson Avenue, Cambridge, CB3 0HE, UK
\and Department of Physics and Astronomy, University College London, Gower Street, London WC1E 6BT, UK 
\and National Astronomical Observatory of Japan, 2-21-1 Osawa, Mitaka, Tokyo 181-8588, Japan
\and INAF-Osservatorio Astrofisico di Arcetri, Largo E. Fermi 5, I-50125, Firenze, Italy
\and European Southern Observatory, Karl-Schwarzschild-Strasse 2, 85748 Garching, Germany
\and Dipartimento di Fisica e Astronomia, Università di Firenze, Via G. Sansone 1, I-50019, Sesto F.no (Firenze), Italy
            }
   \date{}
 
  \abstract
{
The \textit{James Webb Space Telescope} (JWST) is revolutionizing our understanding of black hole formation and growth in the early Universe. However, JWST has also revealed that some of the classical diagnostics, such as the BPT diagrams and X-ray emission, often fail to identify active galactic nuclei (AGN) at high redshift and low metallicity.
Here we present three new rest-frame optical diagnostic diagrams to identify narrow-line Type II AGN, leveraging the \OIIIa auroral line, which has been detected in several JWST spectra. Specifically, we show that high values of the \OIIIa$\bigl/$\Hg ratio provide a sufficient (but not necessary) condition to identify the presence of an AGN, both based on empirical calibrations (using local and high-redshift sources) and a broad range of photoionization models. These diagnostics are able to separate much of the AGN population from Star Forming Galaxies (SFGs). This is because the average energy of AGN's ionizing photons is higher than that of young stars in SFGs, hence AGN can more efficiently heat the gas, therefore boosting the \OIIIa line. We also found independent indications of AGN activity in some high-redshift sources (z>4) that were not previously identified as AGN with the traditional diagnostics diagrams, but that are placed in the AGN region of the diagnostic presented in this work. 
We note that, conversely, low values of \OIIIa$\bigl/$\Hg can be associated either with SFGs or AGN excitation. We note that the fact that strong auroral lines are often associated with AGN does not imply that they cannot be used for direct metallicity measurements (provided that proper ionization corrections are applied), but it does affect the calibration of strong line metallicity diagnostics.
}


   \maketitle
%

\section{Introduction}
\label{sec:intro}

Thanks to the successful launch of the \textit{James Webb Space Telescope} \citep[JWST;][]{Gardner23, Rigby23}, we can now investigate with high resolution and unprecedented sensitivity both the photometric and spectroscopic properties of galaxies up to $z\sim13$ \citep{CurtisLake22, Robertson22}. Within this context, recent studies, exploiting different kinds of JWST data, have revealed a large population of AGN at high redshift \citep{Kocevski23,ubler23,Matthee23,Maiolino23a,Maiolino23c,Greene23,Bogdan23,Goulding23,Kokorev23,Furtak23,Juodzbalis24,Scholtz23b,Chisholm24}, suggesting a higher AGN fraction at $z>3$ than previously expected. For instance, \cite{Yang23}, taking advantage of the JWST-MIRI photometry of the Cosmic Evolution Early Release Science Survey \citep[CEERS;][]{Finkelstein22}, investigated the AGN population using spectral energy distribution (SED) modelling, and found a black-hole accretion rate density (BHARD) at $z>3$ $\sim 0.5$ dex higher than what was expected from previous X-ray studies \citep{vito16,vito18}. Also \citet{Maiolino23c} and \citet{Harikane23}, by selecting broad-line AGN (BLAGN, or TypeI) among NIRSpec spectra \citep{Ferruit23, Jakobsen22}, have found significant AGN excess at $z>4$ with respect to the AGN luminosity functions derived using X-ray data \citep[][]{giallongo19}{}{}. 

The X-ray emission, produced in AGN by inverse Compton scattering of the UV photons coming from the SMBH accretion disc, has been traditionally used to select AGN. Most of the high-z AGN selected using JWST spectroscopy lie in fields covered by some of the deepest X-ray observations ever performed, like the 7Ms Chandra Deep Field South (CDFS) \citep{luo17}, Chandra Deep Field North (CDFN) \citep{xue16} and AEGIS-XD \citep{nandra15}. However, most of these newly discovered high-z AGN lack any X-ray emission \citep[e.g. Mazzolari et al., in prep.,][]{Scholtz23b, ubler23, Maiolino24_X, Kocevski23, Wang24,Juodzbalis2024b}{}{}, with only very few exceptions \citep{Goulding23, Kovacs24, Kocevski23}. This makes the picture even more difficult to interpret and suggests different scenarios such as Compton thick obscuration \citep[][]{gilli22}{}{} or intrinsic X-ray weakness \citep{Simmonds16}.

The difficulties in selecting high-z AGN are even more important for TypeII or narrow-line AGN (NLAGN). Different observational works have already demonstrated that some of the traditional well-known AGN emission-line diagnostic diagrams, like the BPT \citep{Baldwin81} diagram, are no longer effective in the high-z Universe \citep{Maiolino23c,Scholtz23b,Harikane23,Kocevski23,ubler23,Groves06}. Indeed, at high-z, the environments are systematically more metal poor and the stellar populations are generally younger, increasing the ionization parameter ($\rm \log U$) of star-forming galaxies (SFG). The latter effect, on the one hand, makes SFGs move towards the AGN locus on the BPT; on the other hand the lower metallicity of the Narrow Line Region (NLR) of AGN makes its line ratios move towards and overlap with the star formation locus on these traditional diagnostic diagrams.
The detection of high-ionization emission lines (like \NV$\lambda1242$, \NeIV$\lambda2424$, \NeV$\lambda3426$) can still be considered as a safe tracer for the presence of an AGN \citep{Calabro24,Cleri23,Brinchmann23,Chisholm24}, but their detection is still difficult even in the deepest JWST spectroscopic surveys like the JWST Advanced Deep Extragalactic Survey \citep[JADES,][]{bunker23,Scholtz23b}.

To have a clear picture of the AGN properties and demographics at high-z, it is therefore crucial to identify new AGN selection techniques.
Thanks to the large wavelength coverage ($\sim 1-5\mu$m) and sensitivity of JWST it is now possible to have access to a plethora of rest-UV and optical emission lines in high-z galaxies that were not accessible before. Among these lines, multiple JWST spectra have revealed the detection of the \OIIIa line. In the metal poor and highly ionized environments of many high-z galaxies, \OIIIa is the strongest auroral line, i.e. a collisionally-excited emission line generated from higher energy levels compared to the typical nebular lines. Due to its faintness and proximity to the (generally) stronger \Hg line, it has been detected and studied almost only in galaxies at $z<2$ \cite{}, before the launch of JWST. Given its sensitivity to the gas temperature, this line has been extensively used in the local Universe to determine the metallicity of the interstellar medium (ISM) via the so-called ``direct temperature method'', which exploits the possibility of computing the metallicity by fixing the gas electron temperature derived from the ratio between the \OIIIa and the \OIII nebular lines \citep[see e.g.\ review by][]{Maiolino19}. 

However, \cite{Brinchmann23} first noticed that the anomalously high \OIIIa emission in some of the early NIRSpec spectra at z$\sim$8 could be indicating the presence of an AGN. More recently, \citet{ubler23b} have suggested that a strong ratio between this line and the \Hg line could be a possible indicator for the presence of an AGN. In their work, by using JWST/NIRSpec Integral Field Spectroscopy (IFS) observations, they noticed a spatial correspondence between the AGN broad \Hb emission and the \OIIIa line emission. They also mention the possibility that the enhanced \OIIIa emission in correspondence of the AGN could be related to higher ISM temperatures driven by the AGN activity.

An enhancement of the \OIIIa emission in the presence of an AGN is qualitatively expected. Indeed, AGN ionizing photons are more energetic than ionizing photons produced in star forming regions on average, hence they can heat the gas more effectively by depositing a larger amount of energy per unit photon.
In this work, we will explore this scenario more quantitatively, with the ultimate goal of exploring the possibility of selecting AGN using the \OIIIa line both through an observational, empirical approach and via photoionisation models. As a result, we propose three new AGN diagnostic diagrams that can also be effective at high-z.

In Section~\ref{sec:methods} we present the local and high-z observational samples and the set of photoionization models considered to test the validity of the new diagnostic diagrams. In Section~\ref{sec:results}, we describe the new diagnostic diagrams based on the line ratio \OIIIa/\Hg and the demarcation lines that can be used to separate the AGN population from SFGs. Lastly, in Section \ref{sec:discussion}, we provide interpretations on why these diagnostics work and we show indications on the AGN nature of some newly identified AGN using these diagnostics. Then, we also discuss the implications of this work on the metallicity estimates based on the \OIIIa line and on the strong-line metallicity diagnostics.

\section{Methods} \label{sec:methods}
We selected several AGN and SFG samples and two different sets of photoionization models to populate the diagnostic diagrams that we propose to select AGN.
All the observational samples described in the following Sec.~\ref{sec:low-z_obs} and Sec.~\ref{sec:high-z_obs} are listed in Table \ref{table:Samples}.

\subsection{Low redshift samples}\label{sec:low-z_obs}
The sample of low-z SFGs is constituted by both normal SFGs and also by the so-called local analogues of high-z galaxies, i.e. local galaxies that show intrinsic properties similar to those observed in high-z galaxies, in terms of metallicities and emission lines ratios \citep[e.g.][]{izotov18, izotov19, izotov21}.
The Sloan Digital Sky Survey (SDSS) DR7 \citep{Abazajian09} gives the most numerous sample, mostly made of $z<0.7$ galaxies. Starting from the whole catalog provided by MPA/JHU, we considered only sources with a $\rm S/N>5$ in all the lines involved in the diagnostic diagrams we are going to present: \OIIIa, \OIII, \NeIII, \OII, and \Hg. Then we distinguished between AGN and SFGs using the BPT diagram \citep{Baldwin81}, taking the AGN demarcation line provided by \cite{Kewley01} and after considering a further cut in S/N$>5$ also in the \Ha, \Hb and \NII lines. Furthermore, we exclude from the selection AGN classified in the SDSS catalog as BLAGN, since the catalog does not provide separately the NL and BL components of the Balmer lines and our diagnostics rely on the NL emission only. We also correct all the lines for dust attenuation using the \citet{Calzetti00} attenuation law. The final SDSS SFG (AGN) sample considered in this work contains $\sim 2300$ ($\sim 800$) sources.

We also identify several samples of local analogues of high-z galaxies with reported \OIIIa line fluxes. From \cite{yang17} we took the sample of so-called "blueberries", which are 40 dwarf starburst galaxies with small sizes ($<1$kpc), very high ionization (\OIII/\OII$\sim 10-60$) and low metallicities ($\rm 7.1 < 12 + \log(O/H) < 7.8$). They were selected from the SDSS DR12 at $z<0.05$ and followed up spectroscopically using MMT. We also consider 43 "green pea" galaxies from \cite{yang17b}, i.e. nearby SFG with strong \OIII emission line, 2/3 of them also show a strong Ly$\alpha$ emission.

We include in the local analogue sample also a compilation of $\sim 490$ local SFGs with low metallicities from different works: \citet{Izotov06}, \citet{berg12}, \citet{izotov18}, \citet{izotov19}, \citet{Pustilnik20}, \citet{Pustilnik21}, \citet{nakajima22b}. All these samples were selected considering local sources followed up with different telescopes to be sensitive to the faint \OIIIa line, which allowed them to measure the galaxies' metallicity via the so-called direct method. The metallicities of these sources span the range $12+\log(O/H) \sim 6.9-8.9$.

We also include the sample of 165 Extreme Emission Line Galaxies (EELG) reported in \citet{amorin15}, selected from the zCOSMOS survey \citep{lilly07} in the redshift range $0.11<z<0.93$. These are very compact ($r_{50} \sim 1.3$ kpc), low-mass ($\rm M_{*} \sim 10^7 - 10^{10} M_{\odot}$) galaxies characterized by specific star formation rates (sSFR) above the main sequence for star-forming galaxies of the same stellar mass and redshift.

The local AGN sample comprises TypeII AGN showing different features and selected for different aims. The most numerous AGN sample is represented by the SDSS sample mentioned above. We then include the \cite{Dors2020} TypeII AGN sample, selected from SDSS DR7 using multiple AGN diagnostic diagrams to guarantee a high purity and showing a range of metallicities between $\rm 8.0< 12 + \log(O/H) < 9.2$.

The compilation of TypeII AGN includes three sources presented in \cite{Seyfert43}, i.e. NGC1068, NGC1275, and the core of NGC4151. These spectra were obtained at the Mount Wilson Observatory and are among the first galaxies hosting active SMBH ever observed. These three sources clearly show the \OIIIa emission line in their spectra. 

We consider the sources described in \citet{Perna17}, selected from SDSS $z<0.8$ spectra associated with X-ray emission and showing outflows, which can be spectrally decomposed from the narrow and broad line region (BLR) emission. 
We also added both TypeI and TypeII AGN from the S7 survey \citep{dopita15,thomas17}, where 131 nearby AGN were primarily selected because of their radio-detection and then followed-up using the WiFeS instrument at high resolution ($R=7000$), allowing to resolve the faint \OIIIa line. In \citet{thomas17} they provide the narrow components of of the emission lines, already subtracting the broad component, if present.

The 35 AGN selected in \cite{armah21} are TypeII AGN in the local Universe ($z<0.06$) with quite high metallicities (the average metallicity is $\rm 12 + \log(O/H)=8.55$), that the authors select to study the neon to oxygen abundance and its evolution with metallicity.

The last sample of local AGN is taken from the MOSFIRE Deep Evolution Field (MOSDEF) survey \citep{kriek15}, which targeted the rest-frame optical spectra of $\sim 1500$ H-band-selected Galaxies at $1.37 < z <3.8$. We took the full MOSDEF catalogue and selected only those sources with a \OIIIa line with a $\rm S/N>3$, and then we distinguished between AGN and SFG using the BPT diagram. With these S/N cuts, we could identify only 3 sources, all classified as AGN. This indicates the difficulty in detecting such a faint line further than in the local Universe in the pre-JWST era.\\
For all the samples reported above we took dust-corrected fluxes from the works reported in Tab.~\ref{table:Samples}. 

\subsection{High redshift samples} \label{sec:high-z_obs}
The High-z samples of AGN and SFGs only comprise sources observed spectroscopically using JWST, the only instrument with enough resolution, sensitivity and IR-coverage to detect and disentangle the \OIIIa line at $z>3$. 
We considered the full publicly released sample of sources with a medium resolution (MR, $R=1000$) spectrum in the CEERS  \citep[PID: 1345;][]{Finkelstein22} and JADES \citep[PID: 1210;][]{bunker23} surveys. For JADES, we considered the emission line fluxes and the AGN selection described in \citet{Scholtz23b}. Starting from a sample of 110 sources with medium resolution (MR) spectroscopy, a reliable redshift and sufficient wavelength coverage, the authors performed a NLAGN selection using multiple AGN rest-frame UV and optical diagnostic diagrams, finally selecting 28 reliable AGN candidates in the redshift range $1.8<z<9.4$. The diagnostic diagrams involved in the selection include the BPT \citep{Baldwin81}, the VO87 \citep{Veilleux87}, the   CIV$\lambda \lambda 1549,51$/CIII]$\lambda \lambda 1907,1909$ vs CIII]$\lambda \lambda 1907,1909$ / HeII$\lambda 1640$ and the detection of high-ionization emission lines such as [NeV]$\lambda 3420$ [NeIV]$\lambda \lambda 2422,24$, NV$\lambda \lambda 1239,42$. To select NLAGN the authors consider in these diagnostics both demarcation lines already defined in the literature, while in other cases (such as the BPT and VOT) they defined new demarcation lines derived considering the distribution of photoionization models \citep{Gutkin16, Feltre16, Nakajima22}. Of all the initial JADES sources, only 19 show a \OIIIa detection, 7 of which AGN. We followed the same approach for the CEERS program: starting from 313 MR spectra published by the CEERS collaboration, we selected 217 sources with a reliable spectroscopic redshift and we finally selected 45 AGN candidates (Mazzolari et al., in prep). For this work, we considered 32 sources with the \OIIIa line detected among the CEERS sample, 4 of which already classified as AGN. From the same programs, we also collect the BLAGN identified by \citet{Maiolino23c} and \citet{Harikane23} at $z>4.5$, carefully checking, and in case subtracting, the broad component of the \Hg line.

We further consider the three $z\sim8$ galaxies in the galaxy cluster SMACS J0723.3-7327, observed in the Early Release Observations (ERO) JWST programme \citep{Pontoppidan22} and analysed in \citet{Curti23}. Exploiting the \OIIIa line detections in the NIRSpec spectra, the authors measured metallicities ranging from extremely metal-poor ($\rm 12 + \log(O/H) \sim7$) to about one-third solar. Two of them were later identified as AGN in \citet{Brinchmann23} due to the presence of the high-ionization \NeIV\ emission line in one and based on the high ionization parameter ($\rm \log U\sim -1$) of the other. 

The sample of high-z sources presented in \citet{Nakajima23} includes galaxies from three different early JWST programs \citep[CEERS, ERO, GLASS][]{Treu22} at $6<z<9$. We selected only the 10 sources with the \OIIIa line detected, and we marked as BLAGN those that were selected based on their broad \Ha line in \citet{Harikane23}. For the latter, we subtract the broad \Hg component, if present. 

We also include in the high-z AGN sample the BLAGN reported in \citet{ubler23} at $z\sim 5.55$, whose JWST/NIRSpec Integral Field Spectrograph (IFS) observation shows very high ionization lines and low metallicity (1/4 solar). We also include both the components of the dual AGN candidate at $z\sim7.15$ reported in \citet{ubler23b}. For this source, thanks to IFS spectroscopy, the authors were able to study the displacement between the position of the \Hb broad line region (BLR) and the strong \OIII emission line centroid, which were interpreted as the emissions coming from two distinct sources 620pc apart from each other. Interestingly, the authors also found an almost perfect alignment between the \OIIIa emission line peak and the peak of the \Hb BLR, suggesting a correlation between the intensity of the \OIIIa line and the presence of an AGN. 

We also consider the BLAGN reported in \citet{Kokorev23} at $z=8.50$ from JWST UNCOVER Treasury survey \citep{labbe21}, showing a robust \Hb broad component and an unprecedented black hole to host galaxy mass of at least $\sim 30\%$. An even larger black hole to host galaxy mass ratio was more recently found in the $z=6.86$ BLAGN analyzed in \citet{Juodzbalis24}, representing an extreme example of dormant SMBH at high-z. We also considered the $z=7.04$ BLAGN, triply imaged and lensed by the galaxy cluster Abell2744-QSO1, reported by \citet{Furtak23}. This source is probably undergoing a phase of rapid SMBH growth and is also heavily obscured, since the authors measured an $A_V\sim3$ from the Balmer decrement.

Note that in all these TypeI AGN cases (actually most of them type 1.5-1.9) the broad component of the Balmer lines is not detected in H$\gamma$, hence the narrow component of H$\gamma$ can be easily measured.

We added to our high-z sample also the $z=6.1$ extreme SFG reported in \citet{topping24}. Using distict UV transitions the authors found an electron density $\sim 10^4-10^5 \rm cm^{-3}$, a metal poor ionized gas (i.e. $\rm 12 + \log(O/H) = 7.43$), and a $\rm \log U\sim -1$ based on the ratio \OIII/\OII. The authors ultimately classified this source as a SFG, based on the ratios between the rest-UV emission lines, but they also found some indications of possible AGN activity, like broad \Ha component accounting for the $\sim 20\%$ of the line flux.

We finally included in the sample also GN-z11, an exceptionally luminous galaxy at z = 10.6. Its JWST spectrum reveals the presence of an AGN through the detection of semi-forbidden lines tracing very high densities (inconsistent with the ISM but typical of the Broad Line Region), other transitions typical of AGN, fast outflows, ionization cones, and a larger Ly$\alpha$ halo, consistent with those seen in lower redshift quasars \citep{Maiolino23a,Maiolino24_HeII,Scholtz_2023_GN-z11}.

\renewcommand{\arraystretch}{1.1}
\begin{table*}
 \centering
    \input{tabular_sample}
  \caption{Literature data of the low-z and high-z observational samples presented in Sec~\ref{sec:low-z_obs} and Sec.~\ref{sec:high-z_obs} and used to test the effectiveness of the new diagnostic diagrams.}
  \label{table:Samples}
\end{table*}

\subsection{Photoionization models} \label{sec:photmod}
The observational data are compared with the results coming from two different sets of photoionization models. 
The first set of models was initially described in \citet{Gutkin16} (for star formation), and \citet{Feltre16} (for narrow-line AGN emission) and updated with more recent stellar spectra and with a better description of AGN cloud microturbulence \citep{mignoli19,Hirschmann19,vidalgarcia24}. The photoionisation models are built using the \texttt{Cloudy} code (version c13.03) \citep{Ferland13} for star formation and AGN narrow-line regions, assuming a wide range of parameters. In particular they consider gas metallicities ($Z$, where $[O/H]=\log(Z/Z_{\odot})$) in the range $10^{-4}<Z<0.07$, ionization parameters of the ionizing source $-4<\log U<-1$, dust-to-metal ratios ($0.1<\xi<0.5$), hydrogen gas densities ($10^2<(n_H/cm^{-3})<10^4$), different carbon over oxygen abundances (C/O) and two different initial mass functions (IMF) for star formation models. For a complete list of the values of the different parameters we refer to Table 1 of \citet{Feltre16}. These models include predictions for both the AGN and SFG populations; the main feature differentiating between the AGN and SFG models is the ionizing spectrum, i.e. the spectral energy distribution (SED) of the incident ionizing radiation, the first showing a harder radiation field. 

As reported above, the photoionisation models of \citet{Gutkin16} and \citet{Feltre16} cover a wide range of parameters and might actually include physical conditions rarely found in the general populations of galaxies and in the local analogs.
For example, previous theoretical and observational studies have suggested a correlation between the metallicity and ionization parameter in SF regions \citep[e.g.,][]{dopita06, mingozzi20, ji22}, which implies certain combinations of metallicities and ionization parameters should be rarely, if not never, observed in SF regions.
Since the demarcation lines in the new diagrams we propose rely on both the observational data and photoionisation models, we performed the following check to make sure that the part of the models with highly unphysical combinations of parameters does not impact our results.

We performed a likelihood analysis comparing the observed line ratios in our samples of SFG and those predicted by the SFG of \citet{Gutkin16}, in order to limit the inclusion of highly improbable or unphysical combinations of the parameters. This method has been already adopted in making model-based inferences for local galaxies \citep[e.g.,][]{blanc15, mingozzi20, ji22}.


We considered a total of three sets of line ratios, including \OIIIa/\Hg, \OIII/\Hb, and \OIId/\Hg.
The inclusion of collisionally excited lines from different ionization states of oxygen as well as recombination lines of hydrogen helps break the degeneracy between the metallicity and ionization parameter in the models.
Following the formalism of \citet{blanc15}, we calculated the likelihood of each model given each data point, and combined the likelihood with a flat prior in the log space spanned by the metallicity and ionization parameter to obtain the posterior distribution of these two parameters.
From the posteriors, we obtained the weighted average metallicities and ionization parameters for all SFGs.
We then selected a region in the metallicity-ionization parameter space not populated by the posteriors of our sample of SFGs.
We note that while the models of \citet{Gutkin16} and \citet{Feltre16} have a large set of parameters, the primary drivers of the variations in the predicted line ratios are the metallicity and the ionization parameter.
To verify this point, we repeated the above calculation for models with different IMFs, densities, dust-to-gas ratios, and carbon-to-oxygen abundance ratios, which generally resulted in a similar distribution in the metallicity-ionization parameter space.
With these results, we adopted a conservative cut to select realistic SFG models based on the $\rm 16^{th}$ percentile of the inferred distribution in the metallicity-ionization parameter space that produces the highest boundary\footnote{The boundary is determined by the common outer envelopes of SDSS SFGs and local analogs, as these two samples differ significantly in number and occupy different regions in the parameter space.}.

Our final cut removes the region of the SFG parameter space spanned by $\rm \log (Z/Z_\odot)$ and $\rm \log U$ according to the following relations:
\begin{equation}
   \rm  \log U > - 2.8\ \log(Z/Z_\odot) - 3.4,
\end{equation}
and
\begin{equation}
   \rm  \log U > 0.5\ \log(Z/Z_\odot) - 2.245.
\end{equation}
In Fig.~\ref{fig:model_cut}, we plot our fiducial cut together with the inferred distribution of the metallicities and ionization parameters for our selected SFGs based on different sets of models.
The cut basically removes regions having very high ionization parameters and high metallicities at the same time.
From a physical point of view, due to the dependence of the ionizing spectra of young stellar populations and the mechanical feedback from young stellar populations on the metallicity, it is difficult to maintain a very high ionization parameter at a high enough metallicity \citep{dopita06,Carton17}.
While these physically motivated cuts do not impact the location of the AGN photoionization models in the diagnostic diagrams, the parameter space occupied by SFG models is partially reduced. We emphasize that 99.9\% of the considered SFGs lie outside of the excluded parameter space, even assuming different values of $\xi$, $n_H$ and $C/O$.

We also compare the emission line ratios of the observational samples with the models described in \citet{Nakajima22}. Using \texttt{Cloudy}, the authors investigated the emission line ratios of SFG, AGN, Population III stars, and Direct Collapse black holes (DCBHs) using the BPASS stellar population models \citep{Eldridge17} and a range of other parameters (similar to those considered in \citet{Feltre16}) listed in Table 1 of \citet{Nakajima22}. In this work, we only consider the models for SFG and AGN, as the very low metallicity values of the other two models ($Z<10^{-4}$) represent extreme scenarios that are not representative of the sources we want to identify with this analysis. In particular, for these models, the authors considered values of $10^{-5}<Z<10^{-3}$ and values of $\log U$ from -0.5 to -3.5. Since SFG models of \citet{Nakajima22} always cover regions of the diagnostic diagrams already covered by the physically-motivated limited \citet{Gutkin16} and \citet{Feltre16} SFG models, confirming the goodness of our parameters cuts, we did not apply any cut in this case.

\section{Results} \label{sec:results}
In this section, we present the three new diagnostic diagrams proposed to select AGN using the \OIIIa line. In these diagrams we combine the auroral line, sensitive to temperature, with various other lines which are sensitive to other additional properties of the ISM, such as ionization parameter and shape of the ionizing SED, hence can further help to disentangle the source of excitation. Additionally, these lines are accessible with JWST at high redshift, and most of them are also close in wavelength, so little affected by dust reddening.

\subsection{\OIIIa$\bigr/$\Hg vs \OIII$\bigr/$\OII} \label{sec:OH_O32}
The diagnostic presented in Fig.~\ref{fig:diagn_O32} is based on the ratio between the \OIIIa/\Hg compared to \OIII/\OII. Hereafter \OII refers to the sum of \OIId. In the left panel, where we plot the observational samples described in Sec.~\ref{sec:methods}, it is noticeable that the distribution of AGN and SFGs from SDSS are well separated, with AGN occupying a region with higher \OIIIa/\Hg ratios compared to normal SFGs. The distribution of the local analogues in this diagram seems to extend the distribution of the local SDSS galaxies towards higher \OIIIa/\Hg and higher \OIII/\OII, along the diagonal of the plot. On the contrary, the local AGN samples cover a wider area, mostly above the distribution of local analogs and SFGs. 

High-z sources are generally distributed in the upper part of the diagnostic with respect to the local samples. We note that a non-negligible fraction of high-z sources not classified as AGN lies in the region of the diagnostic populated by local AGN, while there are also different high-z AGN falling in the same region covered by local analogues.

On the right panel, we show the distribution of the photoionization models presented in Sec. \ref{sec:photmod} according to the same line ratios. We see here that SFG models are predicted to occupy a limited region of this diagram, whose upper boundary almost perfectly corresponds with the distribution of SDSS SFG and local analogs, while AGN are expected to cover both the region occupied by SFGs and also the part of the diagnostic covered by the local AGN samples. Therefore, SFG models are not able to reach the part of the diagram occupied by local AGN, as expected also from the distribution of the observational samples. The opposite is not true, since AGN can occupy the region populated by SFGs, as observed for some local and high-z AGN. These considerations remain true even considering the full grid of SFG models computed in \citet{Feltre16}, as shown in Fig. ~\ref{fig:all_param_models}.

As we will discuss, to a first approximation, the ratio \OIII/\OII traces the ionization parameter. The fact that, at a given \OIII/\OII, AGN photoionization can reach much higher values of \OIIIa/\Hg than photoionization from hot stars is likely a consequence, on average, of the much higher energy of ionizing photons produced by AGN; therefore, at a given ionization parameter, AGN can heat the gas much more effectively than hot stars.

It is worth noting the position of some high-z sources in this diagnostic. The AGN reported in \citet{Kokorev23} shows an extremely strong \OIIIa line and it is placed in a region that is not covered by any model, which probably suggests a very high electron temperature and $\rm \log U$, as also pointed out in \citet{Kokorev23}. Another significant high-z AGN falling in the AGN-dominated region of this diagnostic is the source corresponding to the BLR centroid in the dual source described in \citet{ubler23b}, where, in the IFS map, the peak of the broad-line emission spatially corresponds to the peak of the \OIIIa line emission. The other component, the \OIII centroid, showing a fainter \OIIIa line emission, is close to the border between the local analogues and the local AGN distribution, together with most of the high-z BLAGN. 
The AGN region is also populated by different AGN selected among the CEERS and JADES MR NIRSPEC spectra, but we also have different sources not selected as AGN but falling in the AGN-dominated region, as pointed out above. We demonstrate in Sec. \ref{sec:discussion} that, from the stack of their spectra, we have an indication that some of these sources may host a BLAGN too.

An interesting source in this diagnostic is the SFG reported in \citet{topping24}. The source shows both strong \OIIIa/\Hg and \OIII/\OII emission line ratios, following the extrapolation of the distribution of the local analogues, but at very high values of the ionization parameters and at low metallicities (see Fig.~\ref{fig:diag_mod_only}), as it is indeed found by the analysis performed in \citet{topping24}. Looking at the photoionization models, that region is not covered by the models of \citet{Gutkin16} and \citet{Feltre16} (not even taking the whole parameters grid), but is instead covered by some AGN models according to \citet{Nakajima22}, potentially suggesting that the main source of ionising radiation in this extreme galaxy is ambiguous.\\

It is worth noting that the effect of dust reddening on the line ratio \OIII/\OII, whose components can suffer a different amount of attenuation due to their distance in wavelengths, would move data points towards the right. This can potentially shift AGNs into the SFG+AGN region under significant reddening, but not the opposite, i.e. galaxies dominated by star formation would not contaminate the AGN-only region.\\


\subsection{\OIIIa$\bigr/$\Hg vs \NeIII$\bigr/$\OII}
The diagnostic presented in Fig. \ref{fig:diagn_Ne3O2} is based on the ratio between the \OIIIa/\Hg compared to \NeIII/\OII. Not surprisingly, the distribution of the local and high-z sources in this diagnostic is very similar to the one described in the previous one. Indeed, \NeIII/\OII correlates closely with \OIII/\OII both in AGN and SFGs, as shown by \citet{Witstok21}, and as expected given that Neon and Oxygen are both $\alpha$-elements and given that Ne$^+$ and O$^+$ have similar ionisation potentials. The disadvantage of \NeIII relative to \OIII is that it is typically much fainter. The advantage of \NeIII is that it is at a shorter wavelength, hence observable at higher redshift (out to z$\sim$12.3 with NIRSpec), and is closer in wavelength to \OII, hence the \NeIII/\OII ratio is much less affected by dust reddening. As for the previous diagram, normal SFGs and local analogs are distributed in the lower-right part of the diagram, in the same region covered by SFG models, while AGN can be distributed over a wider area, including the upper-left part that cannot be reached by any SFG models or samples. As for the diagnostic described in Sec.~\ref{sec:OH_O32}, the separation between the two populations holds even considering the whole grid of the parameter space of the \citet{Gutkin16} star forming models, as presented in Fig.~\ref{fig:diag_mod_only}.

Even if the local analogues are less numerous than in Fig.~\ref{fig:diagn_O32} (simply because there are fewer reported \NeIII fluxes in the literature), their distributions seem to be again an extrapolation of the distribution of local SFGs towards higher $\rm \log U$ and lower metallicities (see Fig.~\ref{fig:diag_mod_only}).

The sources that were clearly in the AGN region in the previous diagnostic are still above the SFG distribution in this diagram, and the sources not identified as AGN but lying in the AGN region of the diagnostic in Fig.~\ref{fig:diagn_O32} are in the same region also in Fig.~\ref{fig:diagn_Ne3O2}. In addition to this latter sample, in Fig.~\ref{fig:diagn_Ne3O2} there are few other sources from the CEERS and JADES samples that were not in the diagnostic of Fig.~\ref{fig:diagn_O32}, because the \OIII line was not available, due to the presence of a detector gap. The AGN region is also populated by three additional sources reported in \citet{Nakajima23} and not classified as AGN in the literature. Their location, however, is still close to the border of the SFG distribution.

The diagnostic diagram presented in this section has some similarity with the so-called `OHNO' diagram that involves the \OIII/\Hb vs \NeIII/\OII line ratios \citep{Backhaus22, Backhaus23, Zeimann15, Cleri23, Trump23,  Feuillet24, Killi23}. The latter is mainly an ionization-sensitive diagram (ionization parameter, shape of the ionizing continuum), but it is also sensitive to metallicity and indeed at high redshift can suffer from low-metallicity galaxy contamination \citep{Scholtz23b}. The O3Hg vs Ne3O2 diagnostic is instead more stable, also at high redshift, as the separation between the AGN and SFG populations is principally based on a different gas temperature, as we will discuss further in Sec.~\ref{sec:discussion}. 

\subsection{\OIIIa$\bigr/$\Hg vs \OIII$\bigr/$\OIIIa}
The third diagnostic is presented in Fig.~\ref{fig:diagn_O33} and is based on the ratio between the \OIIIa/\Hg compared to \OIII/\OIIIa. This diagnostic was reported also in \citet{ubler23b}. 
Looking at the distribution of SDSS SFGs and AGN in the left panel, we note that there is more overlap between the two populations with respect to the previous diagnostics, but there is still a region, characterized by high values of \OIIIa/\Hg and high \OIII/\OIIIa, that is populated only by local AGN. The local analogs distribution seems again to be an extrapolation of the trend of the SDSS SFG distribution. In this diagram, there are very few high-z galaxies that are in the AGN-only region of the diagnostic, while most of them are shifted towards lower values of \OIII/\OIIIa, even in a region that is not covered by the current photoionization models, as it is evident from the right panel. This trend is emblematically represented by the two extreme AGN reported in \citet{Kokorev23} and in \citet{Furtak23}. Given the strength of their \OIIIa line emission, they are placed in the upper-left part of the plot in a region that cannot be reproduced by any model. The same trend is shown by most of the galaxies that were not selected as AGN, but that lie in the AGN region of the two diagnostics presented above. The distribution of the SFG models matches the distribution of the local galaxies, in particular in its boundary with the AGN-only region, while AGN models cover both the region occupied by SFGs and the one populated only by AGN samples. 

In this diagram, assuming the whole parameters' grid of the \citet{Feltre16} models, we would have a complete overlap between the SFG and AGN models. However, the upper left part of the diagnostic would be covered by SFG models characterized by high values of $\rm \log U$ and  high metallicities, a condition not observed in any of our galaxy samples (see Appendix~\ref{sec:app_cutmodels}) and in general in the literature \citep{Kaasinen18, ji22, Grasha22}.

\subsection{Defining the locus of AGN-only objects}\label{sec:demarcation_lines}

For the three diagnostic diagrams described above, we were able to trace clear demarcation lines providing a separation between the AGN-dominated regions and the regions where SFGs and AGN can overlap. To do so, we considered both the photoionization models and the observational sample distributions. 
The demarcation lines for the O3Hg vs O32 diagnostic diagram are defined as follows:
\begin{align}
    &Y=0.55\ X -0.95\ \text{for}\ X>0.84\\
    &Y=0.1\ X -0.57\ \text{for}\ X<0.84.
\end{align}
Similarly, for the O3Hg vs Ne3O2 diagnostic diagram, the demarcation lines are:
\begin{align}
    &Y=0.48\ X -0.42\ \text{for}\ X>-0.07\\
    &Y=0.2\ X -0.44\ \text{for}\ X<-0.07.
\end{align}
Finally, the demarcation line for the O3Hg vs O33 diagnostic diagram is the following:
\begin{equation}
    Y=-1.1\ X +1.47.
\end{equation}
These  demarcation lines ensure that the fraction of contaminating SDSS SFG in the AGN part of the diagnostics is less than 1-2\%, while none of the local analogs lie in the AGN-only part of the three diagnostics.\\
Such a tiny fraction of contaminants can probably be associated with sources that are not identified as AGN by the BPT diagram (used to distinguish between SDSS SFG and AGN in this work), but that host active SMBHs. The presence of AGN contaminants among BPT-selected SFG and local analogs has already been demonstrated by recent works \citep{Svoboda19,Harish23}.\\

Furthermore, it is important to note that the AGN selection based on these demarcation lines is not a necessary condition for a source to be selected as an AGN, but rather a sufficient condition: objects above the demarcation line are AGN-dominated with high confidence, while objects below the demarcation line can either be SFGs or AGN.


\begin{figure*}
	\includegraphics[width=2\columnwidth]{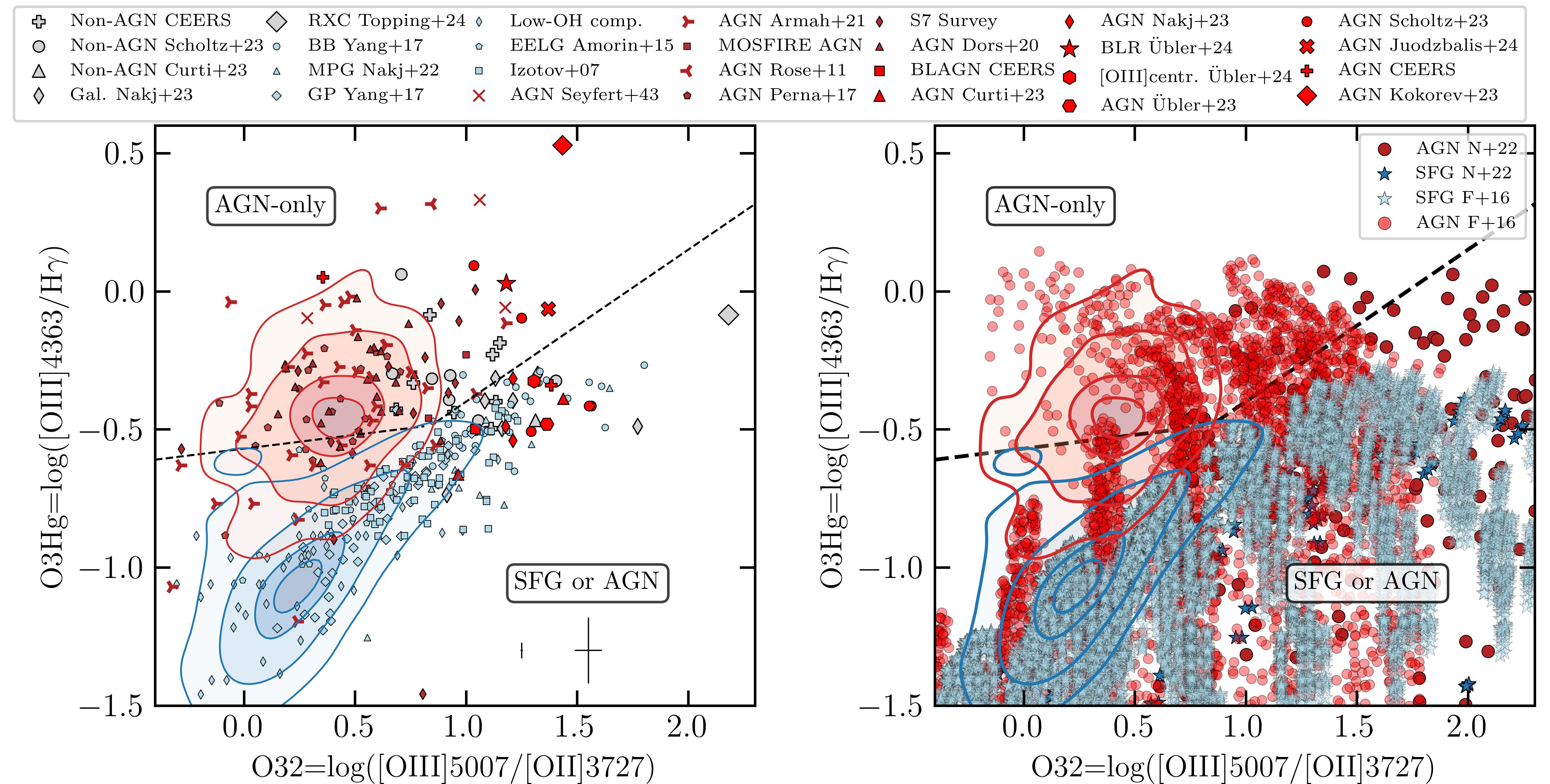}
        \caption{Diagnostic diagram showing the \OIIIa/\Hg vs \OIII/\OII lines ratios. The \OII is the sum of the doublet \OIId. Left panel: plot of all the observational samples described in Sec.~\ref{sec:low-z_obs} and Sec.~\ref{sec:high-z_obs}, with colours and shapes as labelled. Red colours are used for AGN, blue for low-z SFGs, and grey for high-z sources not classified as AGN. The red and blue contours show the distribution at SDSS AGN and SFGs, respectively, including the 90\%, 70\%, 30\%, and 10\% of the populations. Right panel: same plot but showing the AGN and SFG models computed by \citet{Gutkin16}, \citet{Feltre16} and \citet{Nakajima22}, as described in Sec.~\ref{sec:photmod}. The tracks of the AGN and SFG models according to $\log U$ and Z are show in Fig.~\ref{fig:diag_mod_only}. The black dashed line indicates the demarcation defined in Sec.~\ref{sec:demarcation_lines}. Both the distribution of models and of observational samples suggest that the dominant ionizing source in galaxies above the demarcation line is AGN. The error bars reported in the lower-right corner represent the median errors of the low redshift (left) and high redshift (right) samples. The effect of dust attenuation on this diagnostic moves sources towards the right, without contaminating the AGN-only region with SFGs.} 
    \label{fig:diagn_O32}
\end{figure*}

\begin{figure*}
        \includegraphics[width=2\columnwidth]{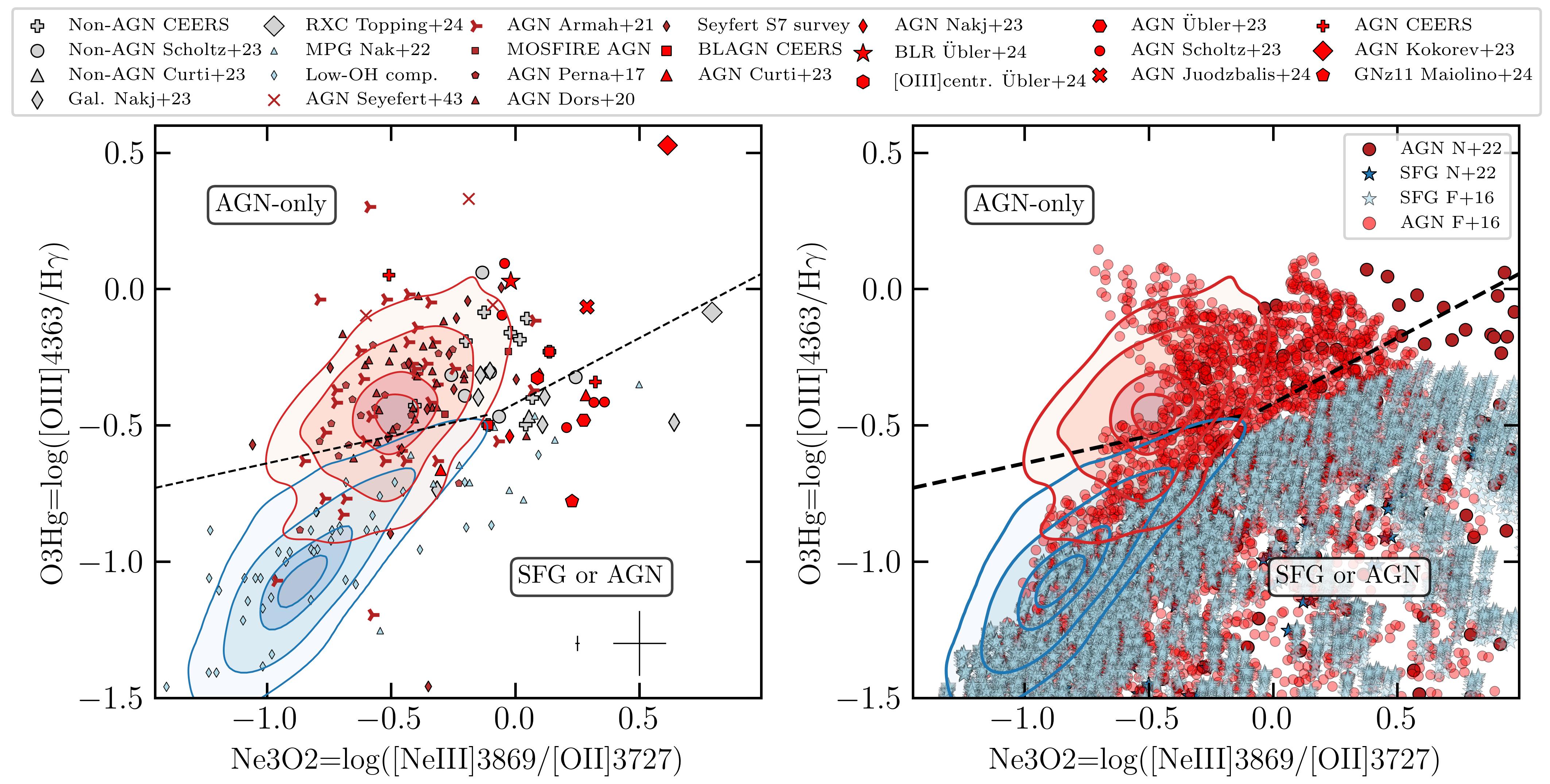}
        \caption{Same as in Fig.~\ref{fig:diagn_O32}, but for the lines ratios \OIIIa/\Hg vs \NeIII/\OII. Based on the distribution of observational samples and models, also this diagnostic diagram identifies a region that can be populated only by AGN.}
    \label{fig:diagn_Ne3O2}
\end{figure*}

\begin{figure*}
        \includegraphics[width=2\columnwidth]{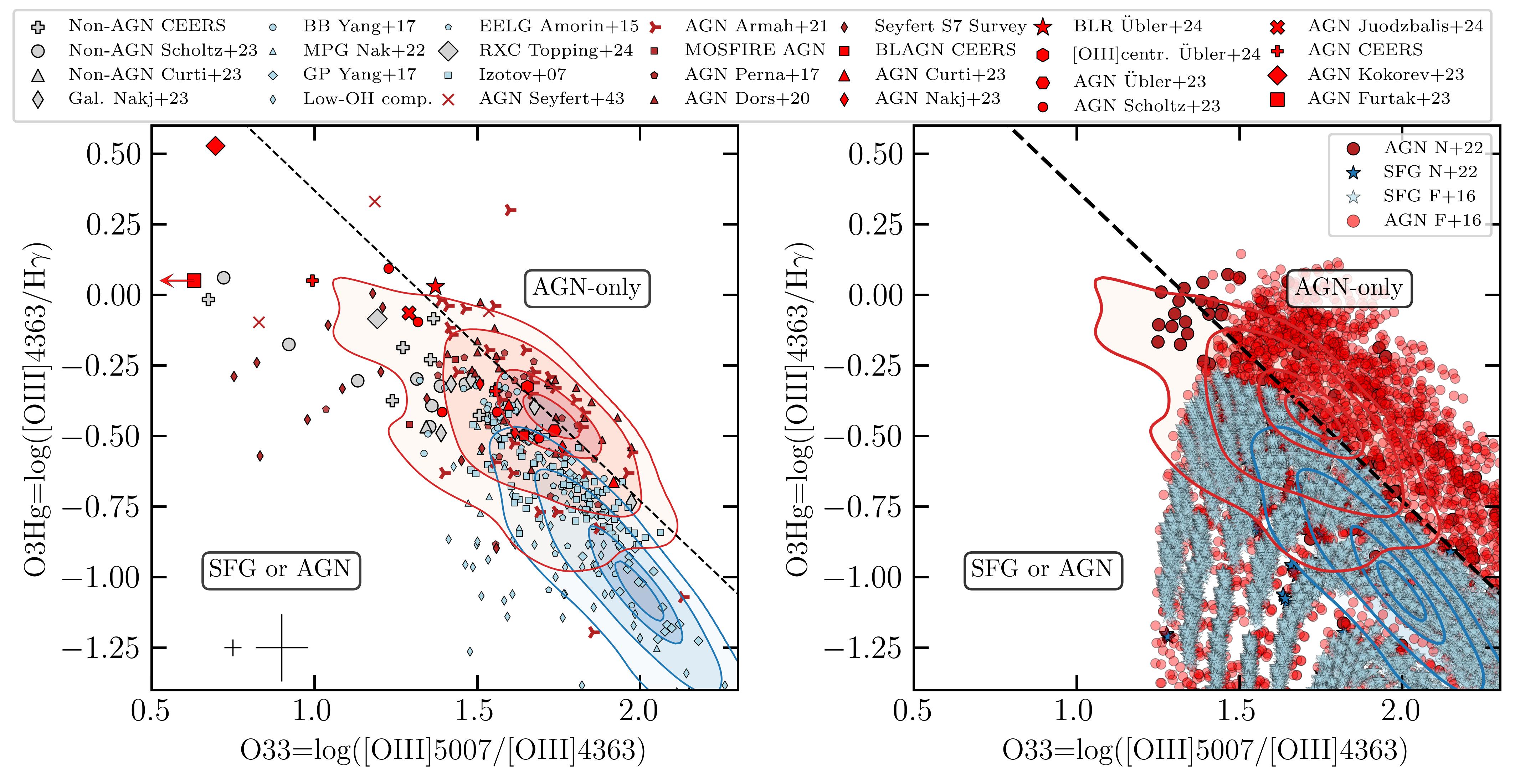}
    \caption{Same as in Fig.~\ref{fig:diagn_O32}, but for the lines ratios \OIIIa / \Hg vs \OIII$\bigl/$\OIIIa. In the left panel, the arrow on the AGN reported in \citet{Furtak23} is for visualization purposes, since it would be located at O3O3$\sim 0$. The cut in SFG models described in Sec.~\ref{sec:photmod} allowed us to identify an AGN-only region of the diagnostic to the right of the black dashed line.}
    \label{fig:diagn_O33}
\end{figure*}

\section{Discussion}\label{sec:discussion}
By comparing the distribution of the observational samples with the distribution of the photoionization models, we can explain why these diagnostic diagrams can separate part of the AGN population from SFGs. In Fig~\ref{fig:diag_mod_only}, we plot the AGN and SFG models in the same diagnostic diagrams presented in Sec.~\ref{sec:results}, but highlighting the variation of $\rm \log U$ and $Z$, i.e. the two main parameters affecting the distribution of the models in these plots. We know that, since they are very local, SDSS AGN and SFG samples are populated mostly by solar metallicity and moderate-to-low ionization parameter sources, as shown also by the models occupying the same region of these samples in Fig~\ref{fig:diag_mod_only}. On the contrary, local analogs are characterized, on average, by lower metallicities and higher ionization parameters (as high-z galaxies). Indeed, in Fig.~\ref{fig:diagn_O32} and Fig.~\ref{fig:diagn_Ne3O2}, they fill the regions that extend from the SDSS SFG towards higher \OIIIa/\Hg and higher \OIII/\OII or \NeIII/ \OII, in close agreement with the trend of photoionization models for higher $\rm \log U$ and $Z$ in the same diagnostics (see Fig.~\ref{fig:diag_mod_only}).

The reason for the presence, in particular in Fig.~\ref{fig:diagn_O32} and Fig.~\ref{fig:diagn_Ne3O2}, of a region that is populated only by AGN, at high \OIIIa/\Hg values, can be explained in the following way. Since the \OIIIa line is a collisionally excited line generated from high energy levels, its intensity relative to \Hg is directly related to the temperature and, secondly, to the metallicity of the ISM and ionization parameter. The main difference between the AGN and SFGs
is the SED of the incident ionizing radiation. For AGN the ionizing radiation is given by the emission of the accretion disk, while for SFGs it originates from young star clusters. The former are characterized by a harder spectrum and, therefore, on average photons are more energetic \citep[see Fig.1 in][]{Feltre16}. Higher energy per ionizing photon, deposited into the ISM, results into a higher effective heating, hence higher electron temperature (hence higher \OIIIa) at a given ionization parameter.

\subsection{Stacking of the AGN candidates}\label{sec:stack}
\begin{figure*}
	\includegraphics[width=1.\columnwidth]{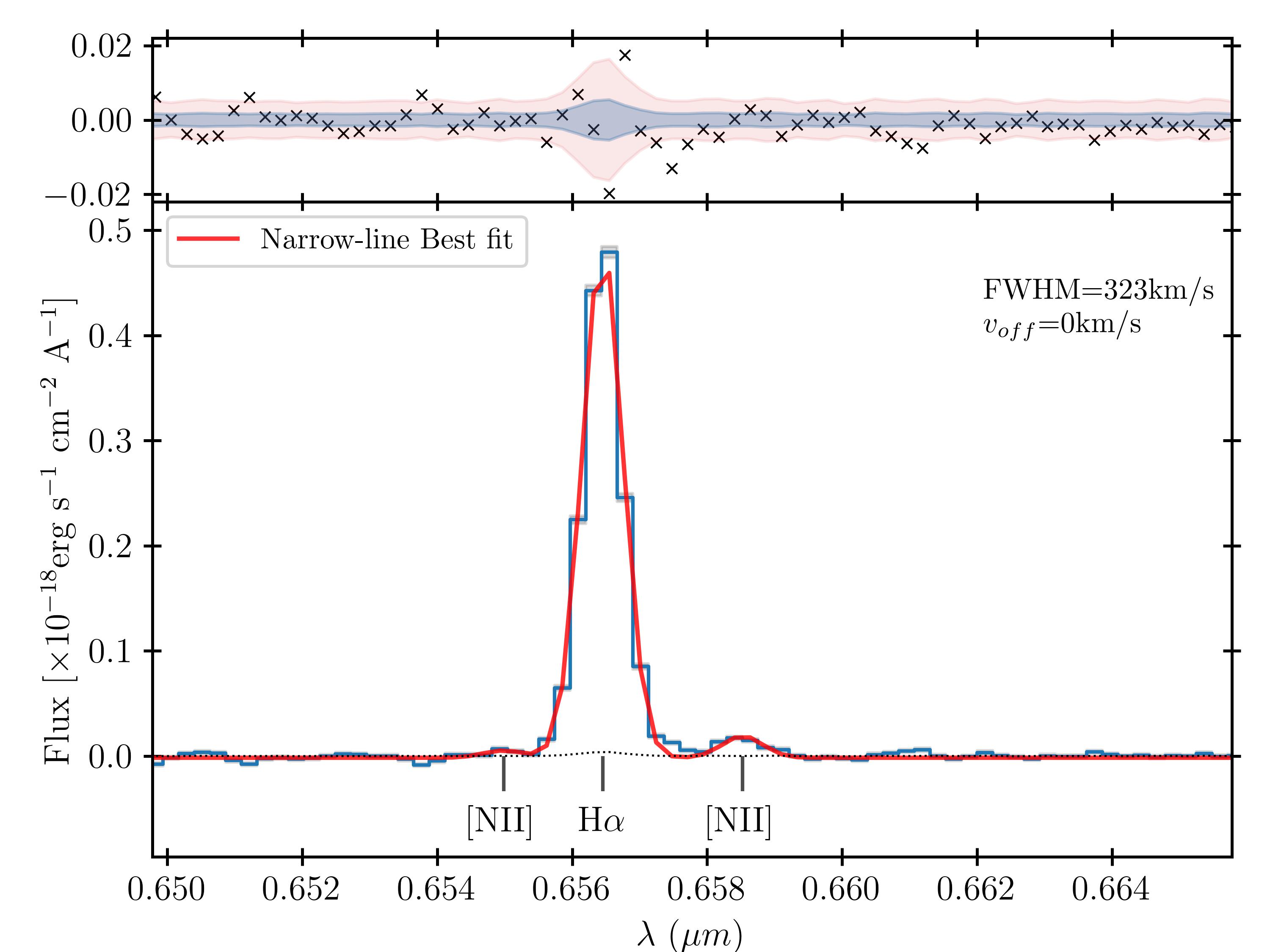}
        \includegraphics[width=1.\columnwidth]{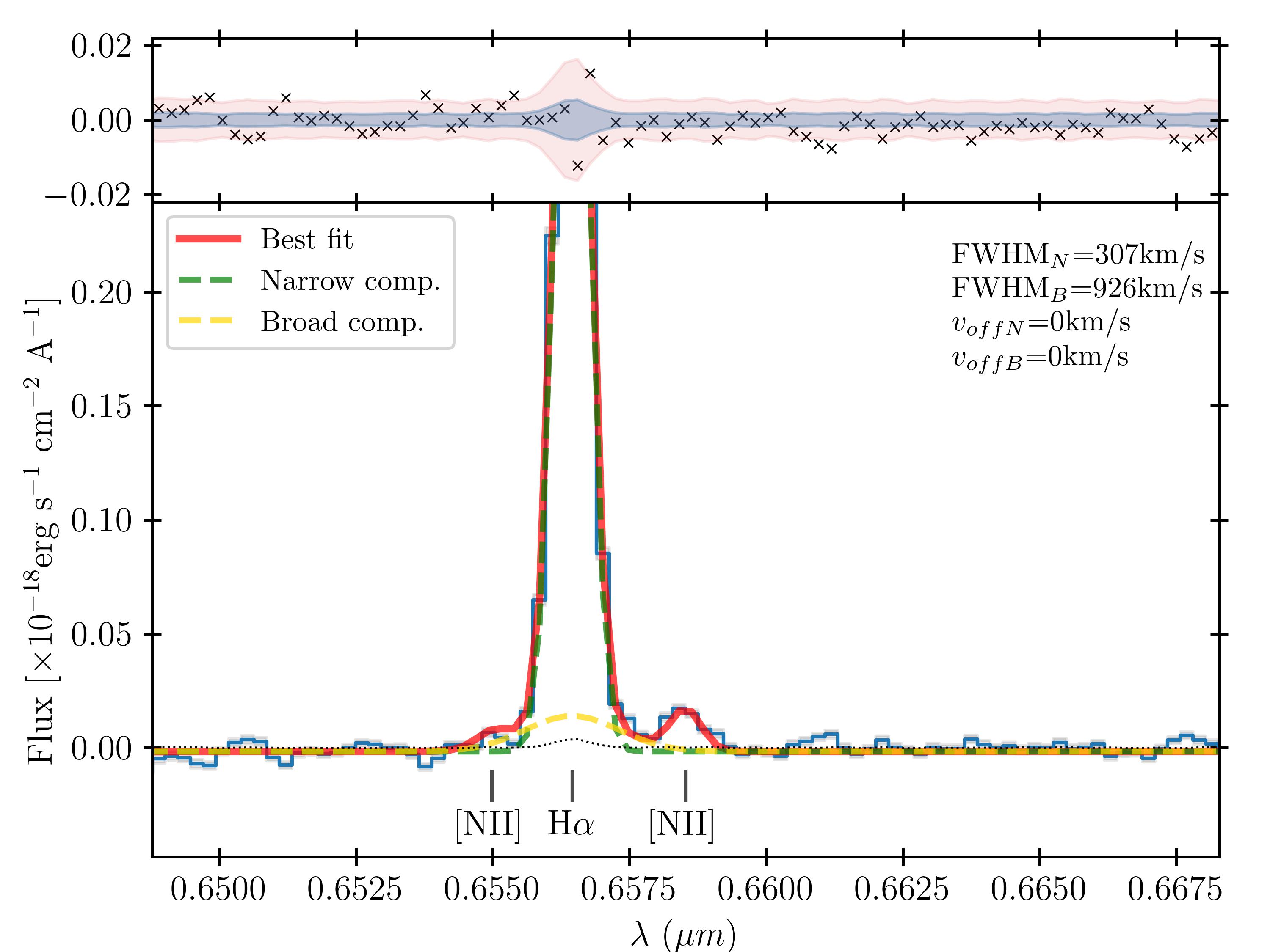}\\
        \includegraphics[width=1.\columnwidth]{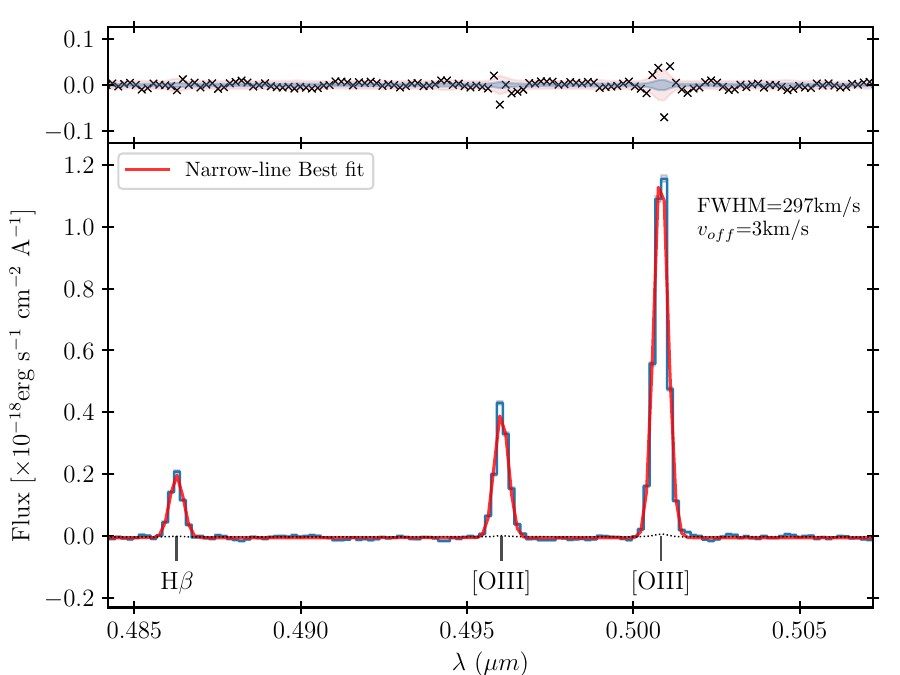}
        \includegraphics[width=1.\columnwidth]{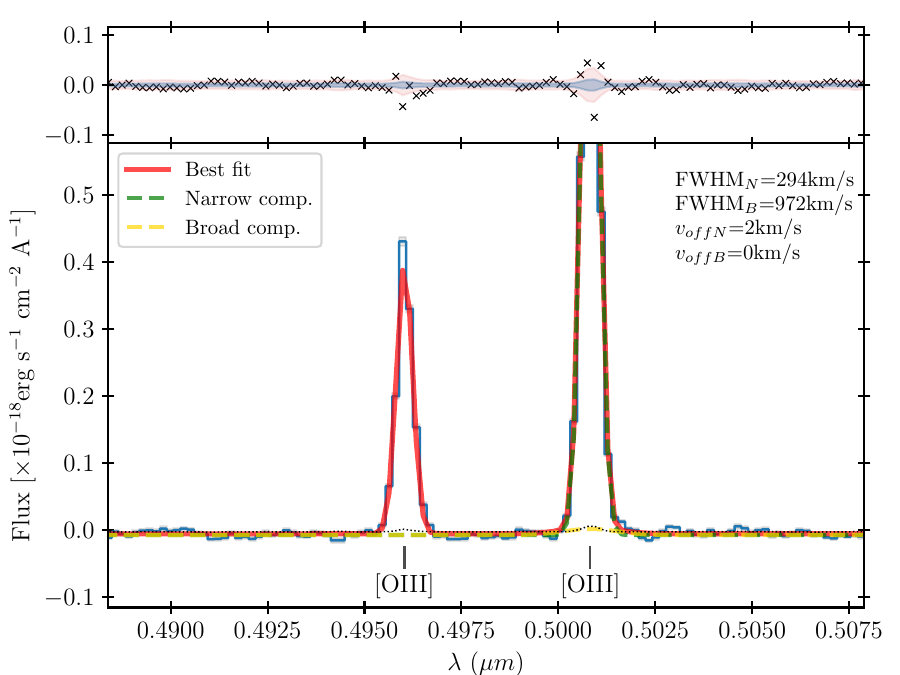}
    \caption{Different fits of the spectrum derived from the stack of the sources not identified as AGN in the literature but lying in the AGN region of the diagnostic diagrams presented in Sec.~\ref{sec:results}. The stacked spectrum in these plots is resampled at the best resolution among those of the single spectra involved (i.e. $\sim 100$km/s), but the results do not change considering the worst resolution (i.e. $\sim 150$km/s).
    Top panels: fit of the \Ha and \NII\, complex with only narrow component (left) and adding a broad component to the \Ha line (right). The global fit is presented in red, while the narrow and broad \Ha components are in green and yellow, respectively. In the upper panel are shown the residuals of the fit compared with the distribution of the $1\sigma$ (blue) and $3\sigma$ (red) errors on the fluxes. On the upper right part of the plots are reported the FWHM and the velocity offset of the different components considered in the fits.
    Lower panels: same as above but for the \Hb and \OIIId doublet complex. In particular, on the right, we added to the fit of the \OIII a broad component with the same FWHM and velocity offset as the broad component of the \Ha, but here it is not required by the fit. }
    \label{fig:stack}
\end{figure*}

To further demonstrate the effectiveness of the diagnostic diagrams proposed in this work we explored potential tracers of AGN activity in those high-z galaxies lying in the AGN region of the diagnostics diagrams in Fig. \ref{fig:diagn_O32} and Fig. \ref{fig:diagn_Ne3O2} but not previously identified as AGN based on the selection performed in \citet{Scholtz23b} and Mazzolari et al., in prep.
Our stacking procedure is as follows. 
We first shift the spectra to the rest-frame and normalise them to the peak of the \Ha line, that is available for all the sources involved. This ensures that the bright sources are not dominating the final stacks. We then resample each of the spectra to the common wavelength grid. We verified that the wavelength grid does not impact our conclusions by repeating the analysis using wavelength bins ranging from the best ($\sim 100$km/s) to the worst ($\sim 150$km/s) resolution. Finally, before stacking the spectra, we fit and subtract the continuum from the rest-frame rebinned spectra. We performed an inverse variance stacking of 15 sources (11 from CEERS and 4 from JADES). \\
\indent We first fit simultaneously the \Ha and \NII wavelength region using only a single Gaussian profile per emission line, with common FWHM and redshift for all three emission lines. We fixed the \NII doublet ratio to be 3. We show the narrow emission line fit in the top left panel of  Fig. \ref{fig:stack}. We detect broad residuals around the \Ha emission line. These broad components can arise from either a broad line region or an outflow.\\
\indent To model the broad wing in the \Ha, we refit the \Ha and \NII with two additional models - BLR model (by fitting one additional broad Gaussian component to \Ha only) and an outflow model by adding broad Gaussian components to both \Ha and \NII with fixed FWHM and centroid. Using the Bayesian Information Criterion (BIC) parameter \citep{Liddle07}, we found that the BLR model is strongly preferred to the narrow-only fit ($\Delta BIC= BIC_{H\alpha \ NL}- BIC_{H\alpha \ BL}= 90$). The broad component has an S/N of 10, making this a solid detection. On the contrary, the fit with the outflow model did not return any broad component in the \NII lines.\\
\indent We then used the same approach with the \Hb and \OIIId complex, to test the possibility that this broad component could be associated with an outflow on larger scales, determining a broad \OIII. In the fit shown in the last panel of Fig.~\ref{fig:stack}, we imposed the broad component of the \OIIId to have the same kinematics as the broad component of the \Ha, i.e. same FWHM and velocity offset. However, we didn't find a significant broad line, and therefore outflow, detection in the \OIIId. In particular, by imposing a broad outflow component to the \OIIId and comparing with the NL-only fit, it gives a $\Delta BIC= BIC_{[OIII]\ NL} - BIC_{[OIII]\ BL}= 4$, meaning that the two models are almost equally preferred, hence a broad \OIII component is not required.\\
\indent The absence of a counterpart in (brighter) \OIII strongly disfavours an outflow origin of the broad component of \Ha. Therefore, the detection of a broad component in the \Ha emission line suggests the presence of an AGN BLR in the spectra of at least some of these galaxies, which is too weak to be detected in the individual spectra. This indicates that galaxies located in the upper region of our diagnostic diagrams, even if originally not classified as AGN, actually may host faint TypeI AGN. This result provides confidence that the diagnostic diagrams presented in this work are effective in identifying AGN above our proposed demarcation lines, although they cannot distinguish between AGN and SFGs below the demarcation line.

\subsection{Ruling out [FeII]$\lambda 4360$ contamination}
We investigate the possibility that the \OIIIa emission line in the high-z sources located in the AGN-only region of the diagnostics could be contaminated by the [FeII]$\lambda 4360$ emission line \citep{Curti17,arellano22}. Most of the high-z \OIIIa detections come from JWST MR spectra, whose resolution is not enough to disentangle the two lines. \citet{Curti17} showed that this contamination is possible, but most likely in high-metallicity galaxies, while our high-z sample is generally metal poor.\\
\indent To investigate the possible iron contamination of the \OIIIa line, we consider the close [FeII]$\lambda 4288$ emission line, which is isolated and originates from the same energy level as the [FeII]$\lambda 4360$. We ran a grid of \texttt{Cloudy} photoionization models over a wide range of metallicities (with $\rm -2 \leq log(Z/Z_\odot) \leq 0.5$) and ionization parameters (with $\rm -4 \leq log(U) \leq -1$), and found that the flux ratio of [FeII]$\lambda 4288$ / [FeII]$\lambda 4360$\footnote{Our calculation takes into account 6 lines from $\rm Fe^+$ blended at 4288 \AA\ and 8 lines from $\rm Fe^+$ blended at 4360 \AA, although the majority of the fluxes is contributed by [FeII]$\lambda 4287.39$ and [FeII]$\lambda 4359.33$.} is roughly a constant of $\sim 1.25$. Therefore, we can use the [FeII]$\lambda 4288$ emission line to reliably constrain the intensity of the [FeII]$\lambda 4360$. We did not detect the [FeII]$\lambda 4288$ line in any of the high-z sources in the AGN-only regions of the diagnostics. Also performing a weighted spectral stacking of all the spectra in the same region, the [FeII]$\lambda 4288$ emission line remains undetected. Therefore we conclude that any possible contribution to the \OIIIa detections given by the [FeII]$\lambda 4360$ emission line is negligible.

\subsection{Impact on metallicity estimates and  strong-line metallicity diagnostics}

The \OIIIa emission line has been widely used in the literature to derive metallicities via the so-called $T_{e}-$method \citep[see ][ for a review]{Maiolino19}, although \cite{Marconi2024} have shown that these metallicities may be biased low. 
The metallicities inferred from the $T_{e}-$method have then been used to calibrate the so-called strong-line metallicity diagnostics, i.e. ratios of  (optical) emission lines typically much stronger than the auroral lines, which can be used to estimate the metallicities on larger samples of galaxies. These calibrations have been inferred both by using local samples of galaxies \citep[e.g.][]{Maiolino008,Bian2018,Curti20} and, especially with JWST, at high redshift \citep{Sanders23,laseter24}.

In most of these past studies care was taken in using only SFGs, or assuming that the gas ionization was dominated by young hot stars. The finding of consistent populations of AGN at high redshift by recent JWST studies, and especially our own finding in this paper that a significant number of strong \OIIIa detections at high redshift are due to AGN heating, may appear as concerning when using this transition to infer the gas metallicity. However, the \OIIIa line (together with \OIII, or \NeIII as a proxy when the latter is not available) provides a measurement of the temperature in the O$^{+2}$ zone, regardless of the nature of the ionizing source. So its reliability remains unaffected in the case of photoionization by AGN \citep[modulo the potential biases discussed in ][ which anyway are a concern also for SFGs]{Marconi2024}. What is obviously important is to properly use this information to infer the metallicity, and in particular by applying the proper ionization correction factors, which are different in the case of AGN and SFGs. In particular, once the abundance of $O^{+2}/H^+$ is derived, one has to estimate the contribution to the abundances from $O^{+3}/H^+$ and  $O^{+}/H^+$. The former is very difficult to assess, even in AGN, as there are no [OIV] strong transitions in the wavelength ranges typically accessible. Yet, \cite{Dors2020} estimate that, even in AGN, the contribution of 
$O^{+3}/H^+$ is typically negligible. The contribution from $O^{+}/H^+$ is estimated from the \OII doublet; however, the key issue is that the temperature of the $O^+$ region is different from the temperature of the $O^{+2}$ region probed by the \OIIIa line. If [OII] auroral lines are not available (as in the vast majority of the cases at high redshift), then one has to assume a relation between $T(O^+)$ and $T(O^{+2})$. This relation has been extensively explored in local SFGs, however it cannot be applied to AGN. Indeed, assuming in AGN the same temperature relation as for SFGs results into a large underestimation of the metallicities \citep{dors15}. This issue is greatly mitigated once AGN-bespoke relations between $T(O^+)$ and $T(O^{+2})$ are adopted \citep{Dors2020}.

In summary, the fact that at high-redshift \OIIIa is boosted by AGN heating does not prevent it from being used for measuring the metallicity in those galaxies, provided that the proper ionization correction factors are adopted and, in particular, the adequate $T(O^+)$ and $T(O^{+2})$ is adopted (in absence of [OII] auroral lines).

However, the presence of AGN ionization, excitation and heating, certainly affects the empirical calibrations of the strong line diagnostics. Indeed, at a given gas temperature, the AGN radiation (with different ionizing shape and, typically, higher ionization parameter) results in a different ionization structure of the gas clouds, and also different collisional excitation rates of the various transitions typically used in the strong line diagnostics ([OIII], [OII], [NeIII], [NII], [SII]). As a consequence, separate and different empirical metallicity calibrations should be inferred for SFGs and AGN host galaxies. This has been attempted in the local universe \citep{Dors2021}. However, the same effort should be undertaken at high redshift, given the large abundance of AGN. Inferring empirical strong-line metallicity calibrations differentiating AGN and SFGs will be the focus of a separate dedicated paper.

\section{Conclusions}\label{sec:conclusions}

While JWST has revealed that some of the classical AGN diagnostics break down at high redshift, it has also opened the opportunity to explore new diagnostics.
In this work, we studied the possibility of selecting AGN using the \OIIIa auroral line, whose detection in large numbers of high-z galaxies has become possible with JWST. In particular, we proposed three new diagnostic diagrams, and three corresponding demarcation lines, that allow to identify a large population of AGN from SFGs, by providing a sufficient (but not necessary) condition to claim the presence of an AGN.

To demonstrate the effectiveness of these diagnostics, we used multiple observational samples of local and high-z SFGs and AGN as well as photoionization models from \citet{Feltre16} and \citet{Nakajima22}.

We specifically proposed the following three diagnostic diagrams:

\begin{itemize}
    \item \OIIIa/H$\gamma$ vs [OIII]5007/[OII]3727. This is the most thoroughly explored diagram from the empirical perspective (it has the largest observational test sample), and can be used with NIRSpec out to z$<$9.4.
    \item \OIIIa/H$\gamma$ vs [NeIII]3869/[OII]3727. This diagram is the least sensitive to dust extinction given the wavelength proximity of both line ratios. Additionally, it can be used with NIRSpec out to z$<$10.9.
    \item \OIIIa/H$\gamma$ vs [OIII]5007/\OIIIa. Among the three, this is the one with the smallest region where AGN can be safely identified, but it is also relatively dust insensitive and requires the detection of only three lines instead of four, making it applicable to larger samples.
\end{itemize}

In each of these cases, we have provided a demarcation line above which (i.e. with higher \OIIIa/H$\gamma$ values) objects can be safely identified as hosting an AGN that dominates the nebular emission lines excitation.

We have illustrated that at least some of few objects falling into the AGN-only region and not previously identified as AGN, do show AGN signatures when stacked. This further supports the effectiveness of the diagnostics. At the same time, we did not find any indication of [FeII]$\lambda4360$ contamination of the \OIIIa emission line in these sources, which could in principle artificially increase the \OIIIa line due to blending effects.

The physics behind these diagnostics is tightly linked to the primary property of AGN, i.e. the hardness of their ionizing spectrum. The average energy of AGN's ionizing photons is much higher than the energy of hot, young stars in SFGs. Therefore, at a given ionizing radiation field intensity, AGN photons are more effective in heating the ionized gas than in SF regions, hence yielding higher temperature, hence higher \OIIIa relative to H$\gamma$ (which normalizes by the overall photoionizing radiation field).

We stress that being above the demarcation line in the proposed diagrams is a sufficient but not necessary condition for an object to be identified as an AGN. Galaxies located below such demarcation lines can be powered by star formation and/or AGN. At the same time, the contamination from SFG is almost completely negligible in the upper part of the diagnostics, as demonstrated by the distribution of low-redshift observational samples, photoionization models. Moreover also the few high-redshift ambigous cases in the AGN-only region turn out to be AGN based on the stacking.

Finally, we note that the fact that strong auroral lines are often associated with AGN does not imply that they cannot be used for direct metallicity measurements (provided that proper ionization corrections are applied), but it does affect the calibration of strong line metallicity diagnostics, calling for new, AGN-specific, calibrations of these diagrams.

\section*{Acknowledgements}
We thank the anonymous referee for useful suggestions which improved the quality of the paper.\\
GM acknowledges useful conversations with Roberto Gilli, Marcella Brusa, Sandro Tacchella, William Baker, Callum Witten, Bartolomeo Trefoloni, Lola Danhaive, Ignas Juodzbalis, Amanda Stoffer, Brian Xing Jiang, and William McClaymont.
FDE, JS and RM acknowledge support by the Science and Technology Facilities Council (STFC), from the ERC Advanced Grant 695671 ``QUENCH'', and by the
UKRI Frontier Research grant RISEandFALL. RM also acknowledges funding from a research professorship from the Royal Society.
H{\"U} gratefully acknowledges support by the Isaac Newton Trust and by the Kavli Foundation through a Newton-Kavli Junior Fellowship.





\bibliographystyle{aa}
\bibliography{literature} 




\appendix

\section{Cut in photoionization models}\label{sec:app_cutmodels} 
\begin{figure*}
	\includegraphics[width=2\columnwidth]{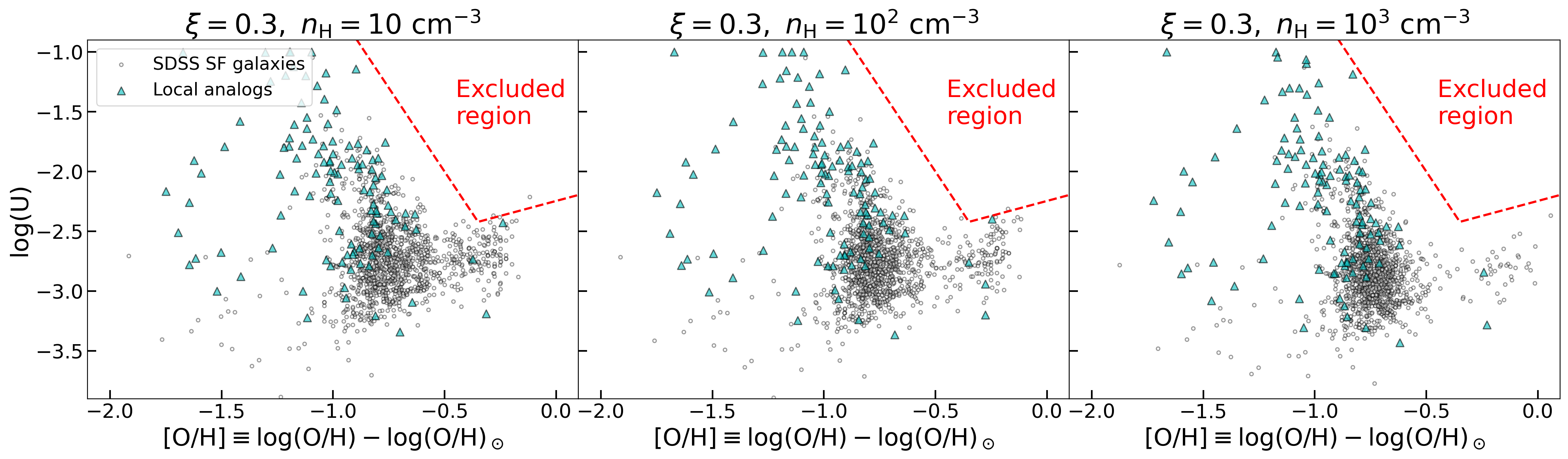}\\
        \vspace{0.5cm}
        \includegraphics[width=2\columnwidth]{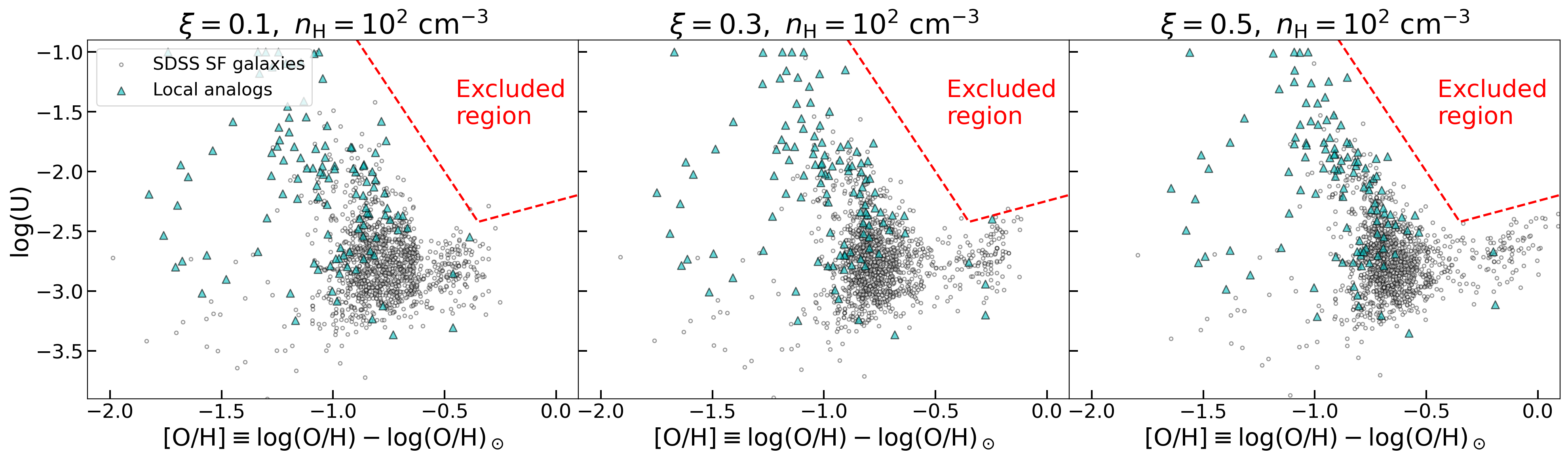}\\
        \vspace{0.5cm}
        \includegraphics[width=2\columnwidth]{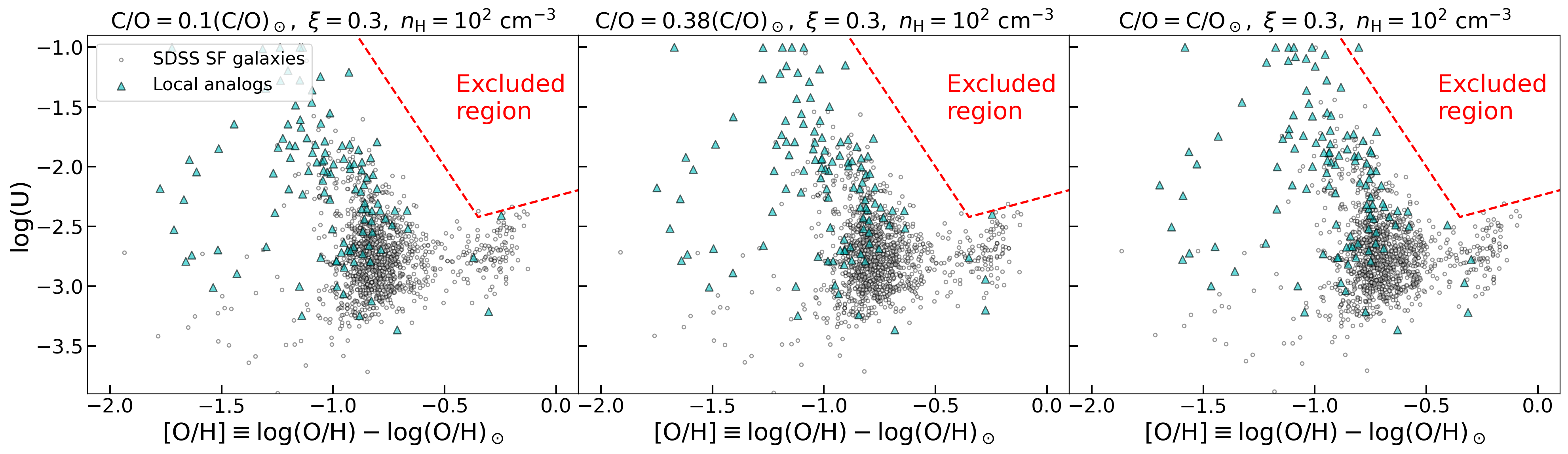}
        \caption{Distribution of the SDSS SFG sample (circles) and local analogues (triangles) in the $\rm \log U$ vs $\rm \log OH$ plane, according to the likelihood procedure with the \citet{Feltre16} models described in Sec.~\ref{sec:photmod}. In each panel, we also show, with a red dashed line, the region of the parameter space that we decided to exclude, a posteriori, from the same photoionization models. The upper panels show the variation of the distribution of the sources by considering models with different electron densities $\rm n_H$, in the middle panels different dust-to-metal ratios $\xi$, and in the bottom panels different $C/O$ abundance. The excluded region is never significantly populated by any source in all the panels. }
    \label{fig:model_cut}
\end{figure*}

In Fig.~\ref{fig:model_cut} we show the distribution of the SDSS SFG and local analogues samples in the $\rm \log U$ vs $\rm \log OH$ plane as derived following the likelihood procedure with the \citet{Gutkin16} models described in Sec.~\ref{sec:photmod}. Even considering different values of the C/O, $\xi$, and $n_H$ parameters in the models, the chosen excluded region contains less than $0.1$\% of the sources (located very close to the borders).

\section{Photoionization models properties}\label{sec:app_models_distrib}
\begin{figure*}
\centering
	\includegraphics[width=1.8\columnwidth]{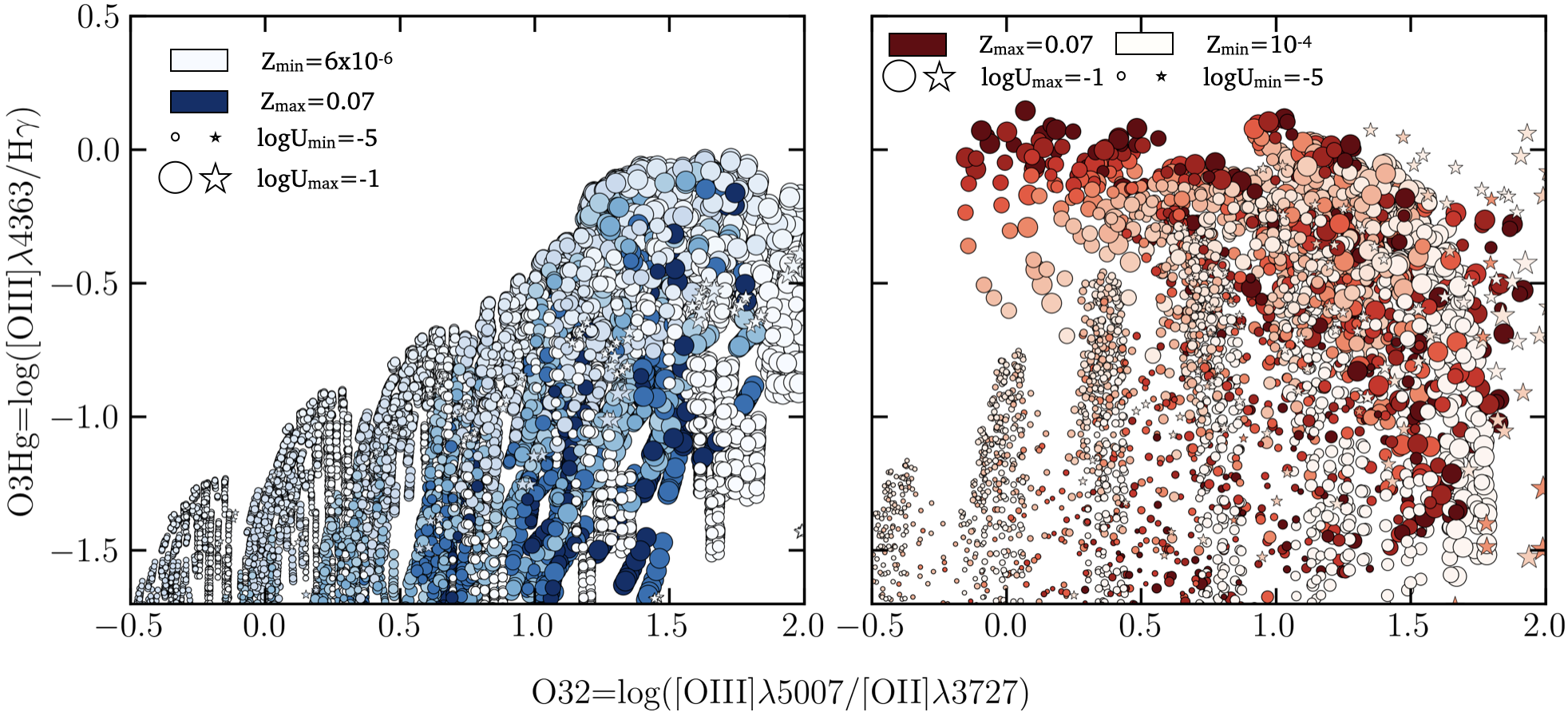}\\
        \vspace{0.5cm}
	\includegraphics[width=1.8\columnwidth]{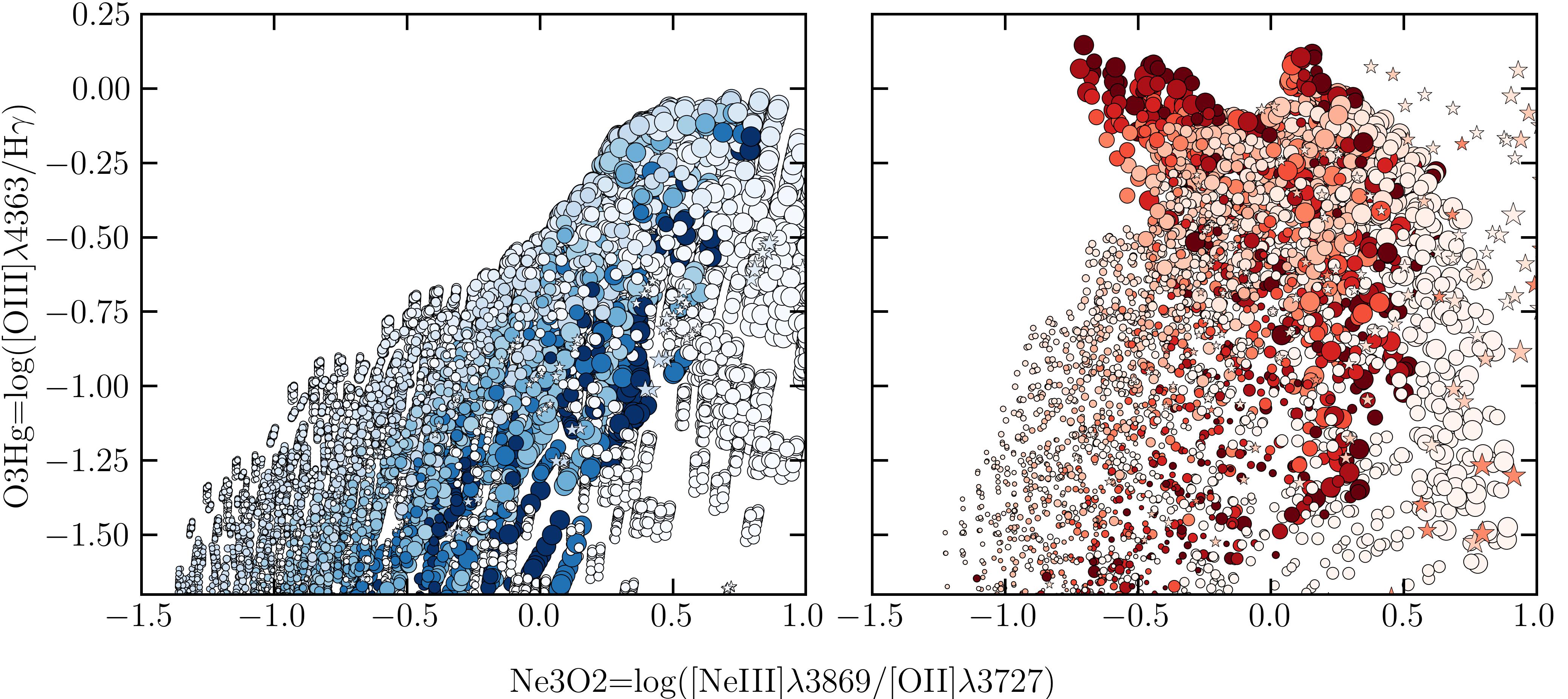}\\
        \vspace{0.5cm}
	\includegraphics[width=1.8\columnwidth]{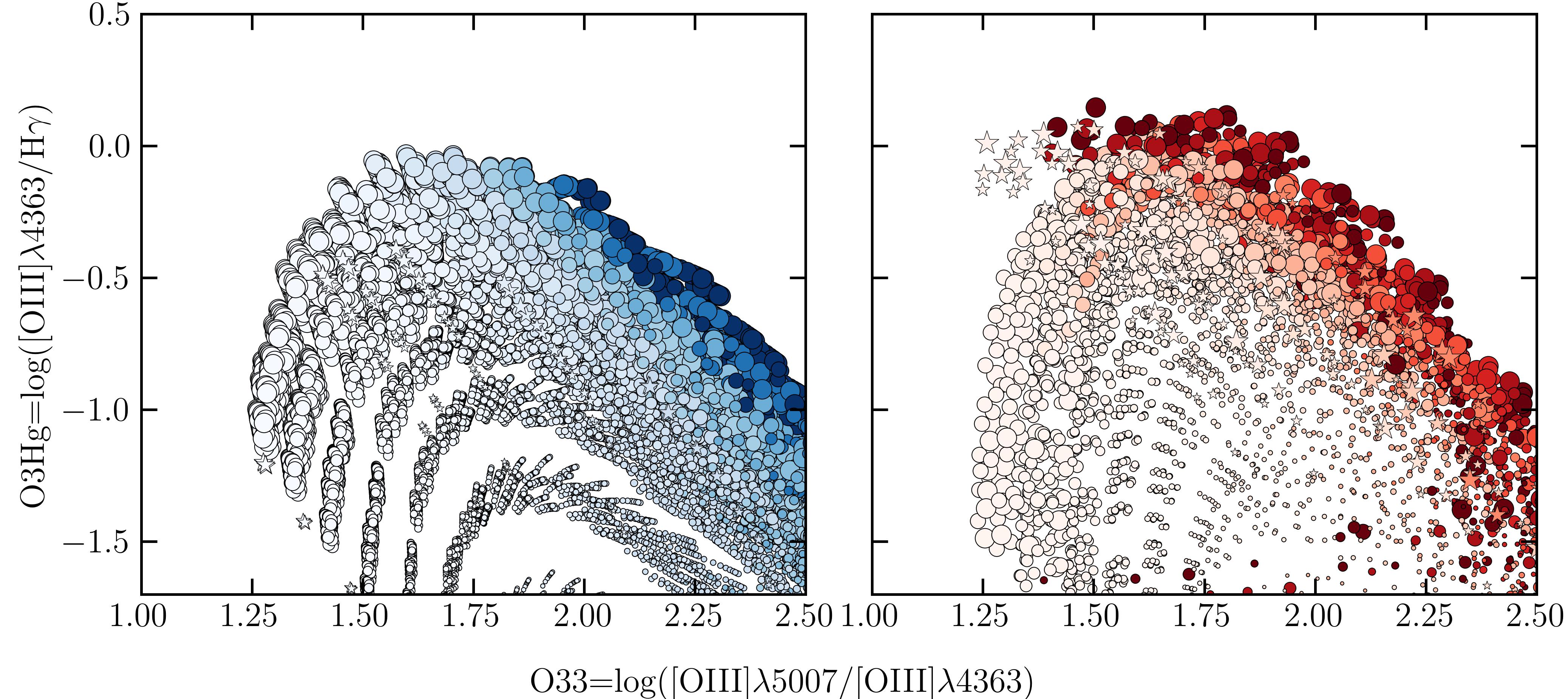}
    \caption{Distribution of the photoionization models computed in \citet{Feltre16} (circles) and \citet{Nakajima22} (stars) with respect to the line ratios reported in Fig.~\ref{fig:diagn_O32} (top panel),Fig.~\ref{fig:diagn_Ne3O2} (central panel), Fig.~\ref{fig:diagn_O33} (bottom panel). The photoionization models for SFG are reported on the right, while the AGN models are on the left. The points are colour-coded according to their metallicity (the stronger the colour, the higher the metallicity), and the marker size depends on the ionization parameter (the larger the marker, the higher the ionization parameter). The maximum and minimum values of $Z$ and $\rm \log U$ for the two classes of models are reported in the top part of the top panels.}
    \label{fig:diag_mod_only}
\end{figure*}

In Fig.~\ref{fig:diag_mod_only} we show the distribution of the SFG and AGN photoionization models from \citet{Gutkin16}, \citet{Feltre16}, and \citet{Nakajima22} in the three diagnostic diagrams presented in Sec.~\ref{sec:results}. In particular, the plots highlight the variation of $\rm \log U$ and $Z$ of the models across the diagnostics.

\section{Diagnostic diagrams including all photoionization models}\label{sec:app_allmodels}
\begin{figure*}
	\includegraphics[width=0.67\columnwidth]{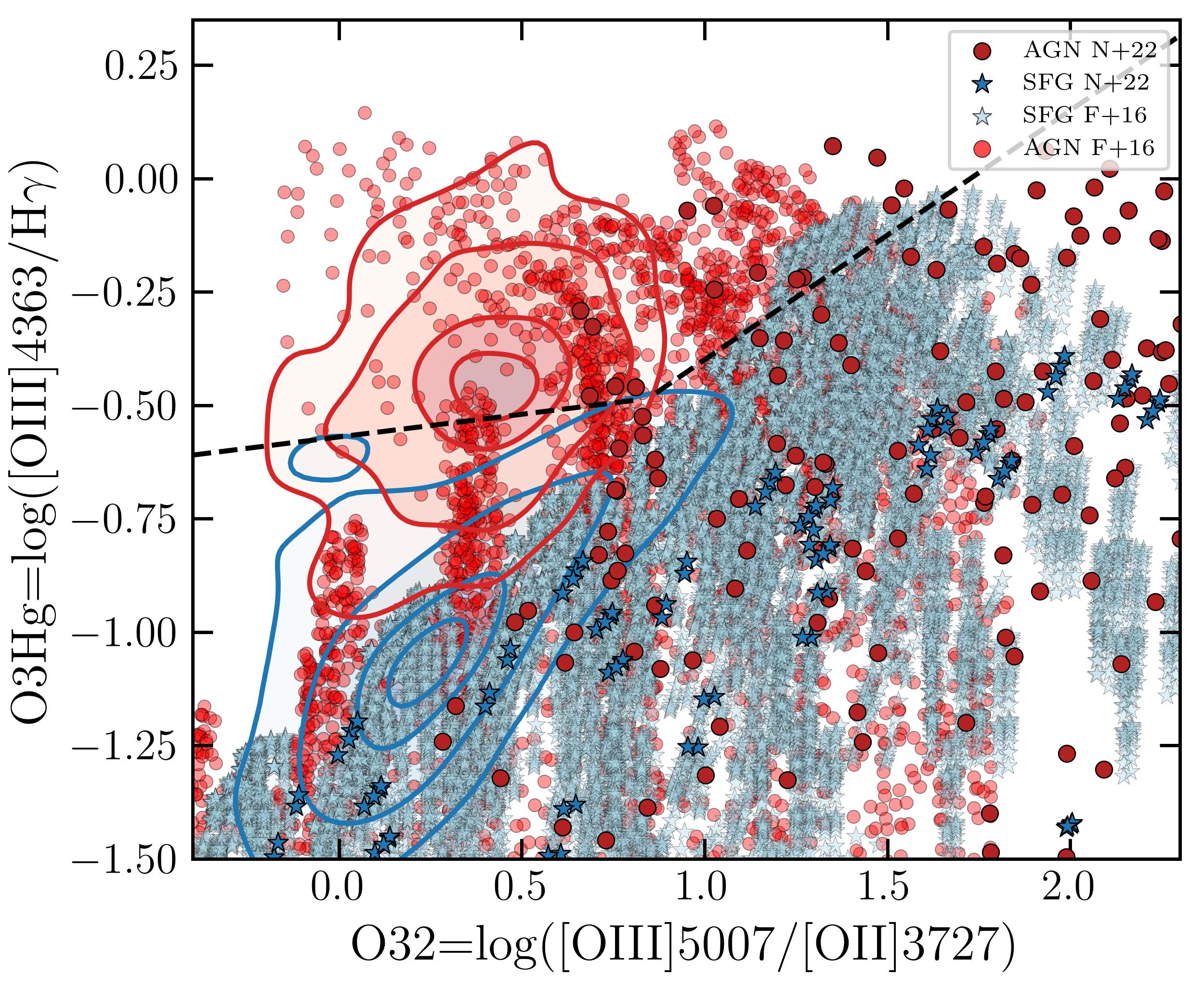}
	\includegraphics[width=0.67\columnwidth]{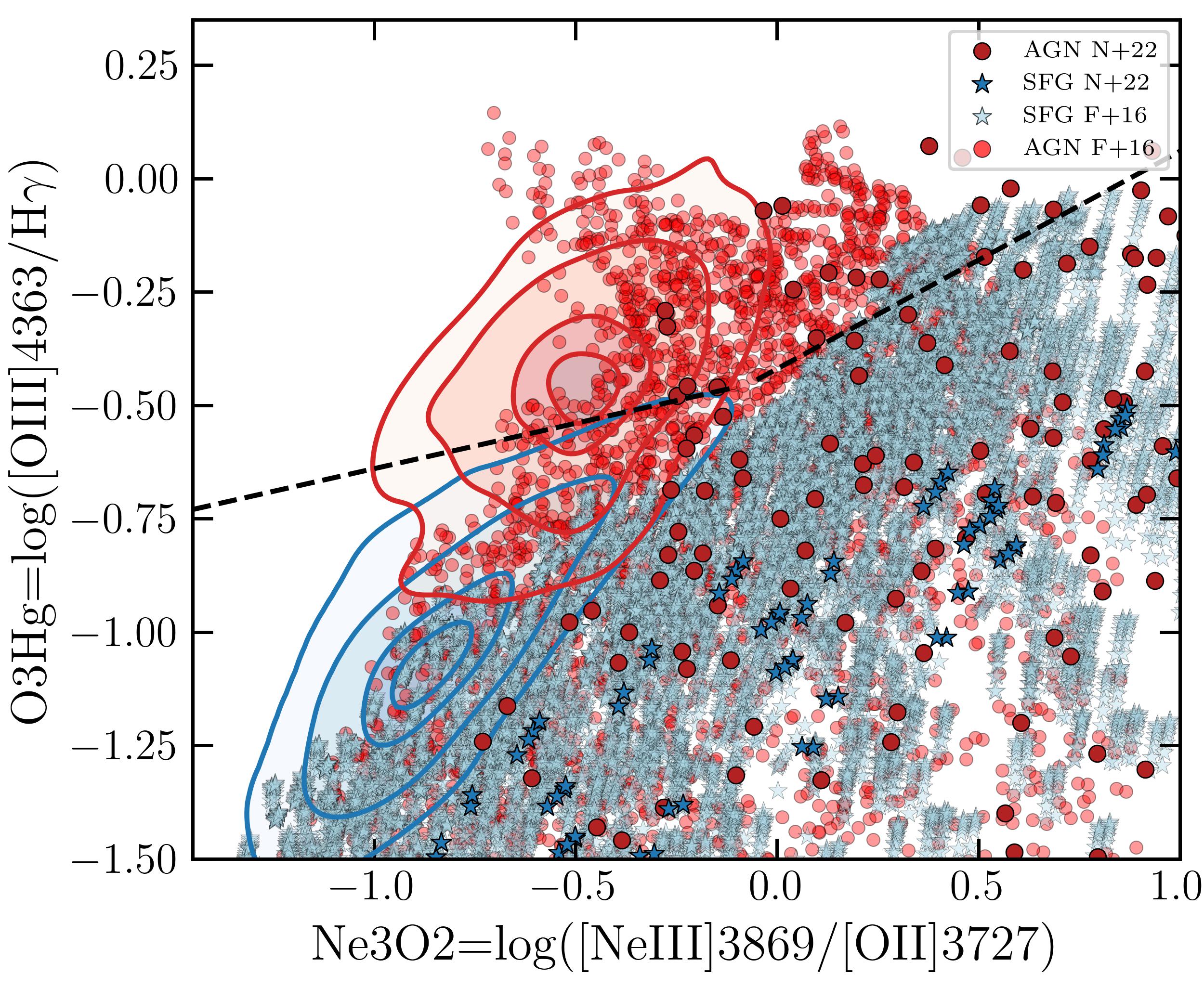}
	\includegraphics[width=0.67\columnwidth]{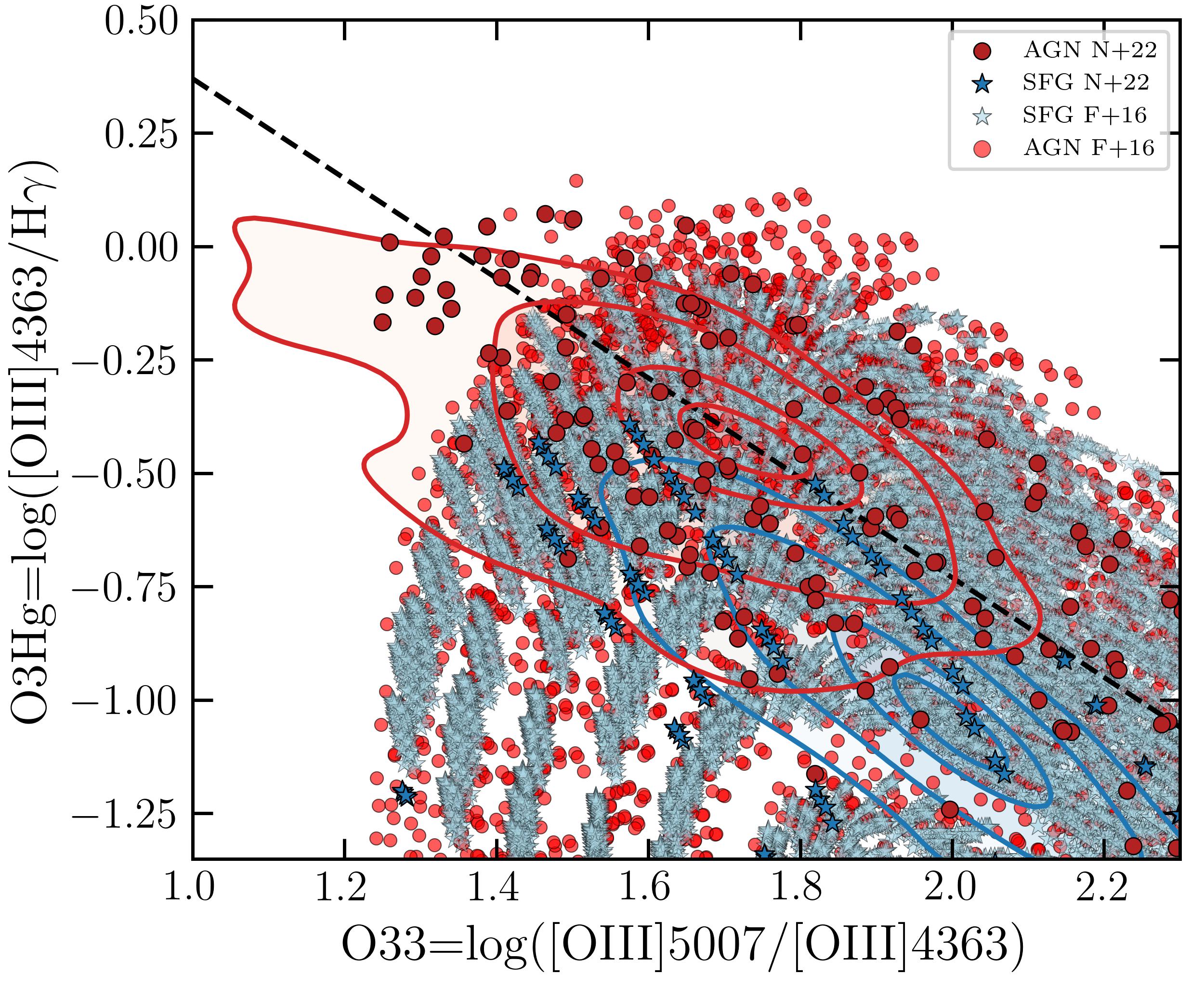}
    \caption{From left to right, same plots reported in the right panels of Fig.~\ref{fig:diagn_O32}, Fig.~\ref{fig:diagn_Ne3O2} and Fig.~\ref{fig:diagn_O33}, but showing all the models computed in \citet{Gutkin16}, \citet{Feltre16} and \citet{Nakajima22}, i.e. without the cut described in Sec.~\ref{sec:photmod}. We note that the demarcation lines for the O3Hg-O32 and O3Hg-Ne3O2 still hold, while there is an almost complete superposition between SFG and AGN models in the O3Hg-O33 diagnostic diagram.}
    \label{fig:all_param_models}
\end{figure*}

The demarcation lines presented in Sec.~\ref{sec:demarcation_lines} were defined considering both the distribution of the observational samples and of photoionization models with the cut described in Sec.~\ref{sec:photmod} and shown in Fig.~\ref{fig:model_cut}. If we consider the entire grid of parameters for the models of \citet{Gutkin16} and \citet{Feltre16}, we obtain the distributions shown in Fig.~\ref{fig:all_param_models}. In this case, the demarcation lines for the O3Hg-O32 and O3Hg-Ne3O2 diagnostic plots still hold, except for a very small fraction of SFG models populating the AGN-only part of the two diagnostics. On the contrary, in the O3Hg-O33 diagnostic plot, there is an almost complete superposition between the SFG and AGN models. Even if this superposition is determined by highly unlikely combinations of parameters in the models, we cannot completely rule out the possibility that the AGN-only region of the O3Hg-O33 diagnostic diagram could actually be partially populated also by SFG. 


\end{document}

%% file: tabular_sample.tex
\begin{tabular}{@{}lcccc@{}} 
\hline 
\hline 
Sample/References & Type  & N included & Redshift\\
 \\
\hline 
SDSS DR7 \cite{Abazajian09} & SFG \& AGN & 3188 (2377/811) & $z_{med}\sim 0.1$ & \\
\cite{yang17} & Blueberries Galaxies & 40 & $z<0.05$ & \\
\cite{yang17b} & Green Pea Galaxies & 43 & $z<0.05$ & \\
Low-OH compilation & Low-metallicities SFG & 489 & $z<0.1$ & \\
\cite{amorin15} & Extreme Emission Line Galaxies & 165 & $z<1$ & \\
\cite{Seyfert43} & AGN & 3 & $z\sim 0$ & \\
\cite{Dors2020} & AGN & 26 & $z<0.05$ & \\
\cite{Perna17} & AGN & 44 & $z<0.8$ & \\
S7 Survey \cite{dopita15} & AGN & 313 & $z<0.1$ & \\
\cite{armah21} & AGN & 35 & $z<0.06$ & \\
\cite{kriek15} & AGN & 3 & $z\sim 2.5$ & \\
\hline
CEERS (R1000) \cite{Finkelstein22} & SFG \& AGN & 32 (28/4) & $2.3<z<8.8$ & \\
JADES (R1000) \cite{bunker23} & SFG \& AGN & 19 (12/7) & $0.66<z<9.43$ & \\
\cite{Curti23b} & SFG \& AGN & 3 & $z\sim 8$ & \\
\cite{Nakajima23} & SFG \& AGN & 10 & $6<z< 9$ & \\
\cite{topping24} & SFG & 1 & $6.1$ & \\
\cite{ubler23} & AGN & 1 & $5.55$ & \\
\cite{ubler23b} & AGN & 2 & $7.15$ & \\
\cite{Juodzbalis24} & AGN & 1 & $6.86$ & \\
\cite{Kokorev23} & AGN & 1 & $8.50$ & \\
\cite{Furtak23} & AGN & 1 & $7.04$ & \\
\cite{Maiolino23a} & AGN & 1 & $10.6$ & \\

\hline
\end{tabular} 